\documentclass[preprint]{aastex}
\usepackage{rotating}

\shorttitle{Expanding Photosphere Method}
\shortauthors{Jones et al.}

\begin{document}

\title{Distance determination to 12 Type II Supernovae \\
    using the Expanding Photosphere Method}

\author{M. I. Jones\altaffilmark{1}, M. Hamuy\altaffilmark{1},
P. Lira\altaffilmark{1}, J. Maza\altaffilmark{1},
A. Clocchiatti\altaffilmark{2}, M. Phillips\altaffilmark{3}, \\
N. Morrell\altaffilmark{3}, M. Roth\altaffilmark{3},
N. B. Suntzeff\altaffilmark{4}, T. Matheson\altaffilmark{5},
A. V. Filippenko\altaffilmark{6}, \\ R. J. Foley\altaffilmark{6}, and
D. C. Leonard\altaffilmark{7}}

\altaffiltext{1}{Departamento de Astronom\'ia, Universidad de Chile.}
\altaffiltext{2}{Departamento de Astronom\'ia y Astrof\'isica,
Pontificia Universidad Cat\'olica de Chile.}
\altaffiltext{3}{Las Campanas Observatory, Carnegie Observatories, 
La Serena, Chile.}
\altaffiltext{4}{Texas A\&M University, Physics Department, College 
Station, TX.}
\altaffiltext{5}{National Optical Astronomy Observatory, 950 North 
Cherry Avenue, Tucson, AZ 85719-4933.}
\altaffiltext{6}{Department of Astronomy, University of California, 
Berkeley, CA 94720-3411.}
\altaffiltext{7}{Department of Astronomy, San Diego State University, 
San Diego, CA 92182.}

\begin{abstract}

We use early-time photometry and spectroscopy of 12 Type II plateau
supernovae (SNe~IIP) to derive their distances using the expanding
photosphere method (EPM). We perform this study using two sets of
Type II supernova (SN~II) atmosphere models, three filter subsets
($\{BV\}$, $\{BVI\}$, $\{VI\}$), and two methods for the host-galaxy
extinction, which leads to 12 Hubble diagrams. We find that
systematic differences in the atmosphere models lead to $\sim\,$50$\%$
differences in the EPM distances and to a value of ${\rm H_0}$ between 
52 and 101 ${\rm km~s^{-1}~Mpc^{-1}}$. Using the $\{VI\}$ filter subset 
we obtain the lowest dispersion in the Hubble diagram, \mbox{${\rm
\sigma_{\mu}\,=\,0.32}$ mag}. We also apply the EPM analysis to the
well-observed SN IIP 1999em. With the $\{VI\}$ filter subset we
derive a distance ranging from 9.3 $\pm$ 0.5 Mpc to 13.9 $\pm$ 1.4 Mpc
depending on the atmosphere model employed.

\end{abstract}


\section{Introduction}

Type II supernovae (SNe II) are understood as the result of the final
gravitational collapse of massive stars ($M > 8$~M$_{\sun}$) that at
the moment of the explosion have most of their hydrogen envelope
intact. The energy released in the explosion is typically
$\sim\,10^{53}$ erg (mainly radiated in the form of neutrinos), and
the luminosity of the SN during the first few months after explosion
can be comparable to the total luminosity of its host galaxy. These
objects have been classified based on their light curves into Type IIP
(plateau) and Type IIL (linear) \citep[e.g.,][]{Pat94}. The former
present a nearly constant optical luminosity during the photospheric
phase ($\sim\,$100 days after explosion), while the latter show a slow
decline in luminosity during that phase. However, there are some SN II
events, such as the SN 1987A, that show peculiar photometric
properties. Also, studies of SN~II spectra have revealed the
existence of a subclass, characterized by the presence of narrow
spectral lines, called SNe IIn \citep{Sch90,fil91a,fil91b}, which are most 
likely originated from the interaction of the SN ejecta with pre-existing 
circumstellar material; see \citet{fil97} for a general review of SN spectra.
\newline \indent
Due to their high intrinsic luminosities, SNe~II have great potential as
extragalactic distance indicators. To date, several methods have been
proposed to derive distances to SNe II, but two are the most commonly
used: the expanding photosphere method (EPM) \citep{Kir74} and the
standardized candle method (SCM) \citep{Ham02}. The former is a
geometrical technique that relates the photospheric radius and the
angular radius of a SN in order to derive its distance, and has been
applied to several SNe to derive the Hubble constant
\citep[e.g.,][]{SKE92}. The EPM is independent of the extragalactic
distance ladder, and therefore does not need any external calibration.
The SCM is based on the observed relation between expansion velocity
and luminosity of SNe~IIP. Recently, this method has been applied to a
sample of high-redshift SNe \citep{Nug06}. Other methods have also
been used to determine distances to SNe~II, such as the
spectral-fitting expanding atmosphere method (SEAM) \citep{Bar04} and
the plateau-tail relation proposed by \citet{Nad03}.
\newline \indent
In this work we apply the EPM using early spectroscopy and photometry
of 12 SNe~IIP in order to derive their distances. We apply the method
using two sets of SN~II atmosphere models \citep{E96,D05a}, three
filter subsets ($\{BV\},\, \{BVI\},\, \{VI\}$), and two methods for the
host-galaxy extinction. The different combinations lead to 12 Hubble
diagrams. Section 2 of this paper describes the photometric and
spectroscopic observations. In \S~3, the EPM is presented, and we
apply it to 12 SNe~IIP. The results are discussed in \S~4. We compare
our EPM distances with results from other methods and with previous
EPM analyses. We also discuss the error analysis and the effect of
reddening on the EPM distances. We show 12 Hubble diagrams and the
corresponding Hubble constants, and we propose an external calibration
for the EPM. Finally, we summarize our conclusions in \S~5.

\section{Observations}

In this work we use photometry and spectroscopy from four SN follow-up
programs: the Cerro Tololo supernova program (1986--1996), the Cal\'an/Tololo
supernova survey (CT; 1990--1993), the Supernova Optical and Infrared Survey
(SOIRS; 1999--2000) and the Carnegie Type II Supernova Program (CATS;
2002-2003). During these programs optical (and some IR) photometry
and spectroscopy were obtained for nearly 100 SNe, 51 of which belong
to the Type II class. All of the optical data have already been
reduced and will soon be published \citep{Ham08}. We
also complemented our dataset with some spectra from various
coauthors of this paper.

\subsection{Photometry}

\begin{figure}
\plotone{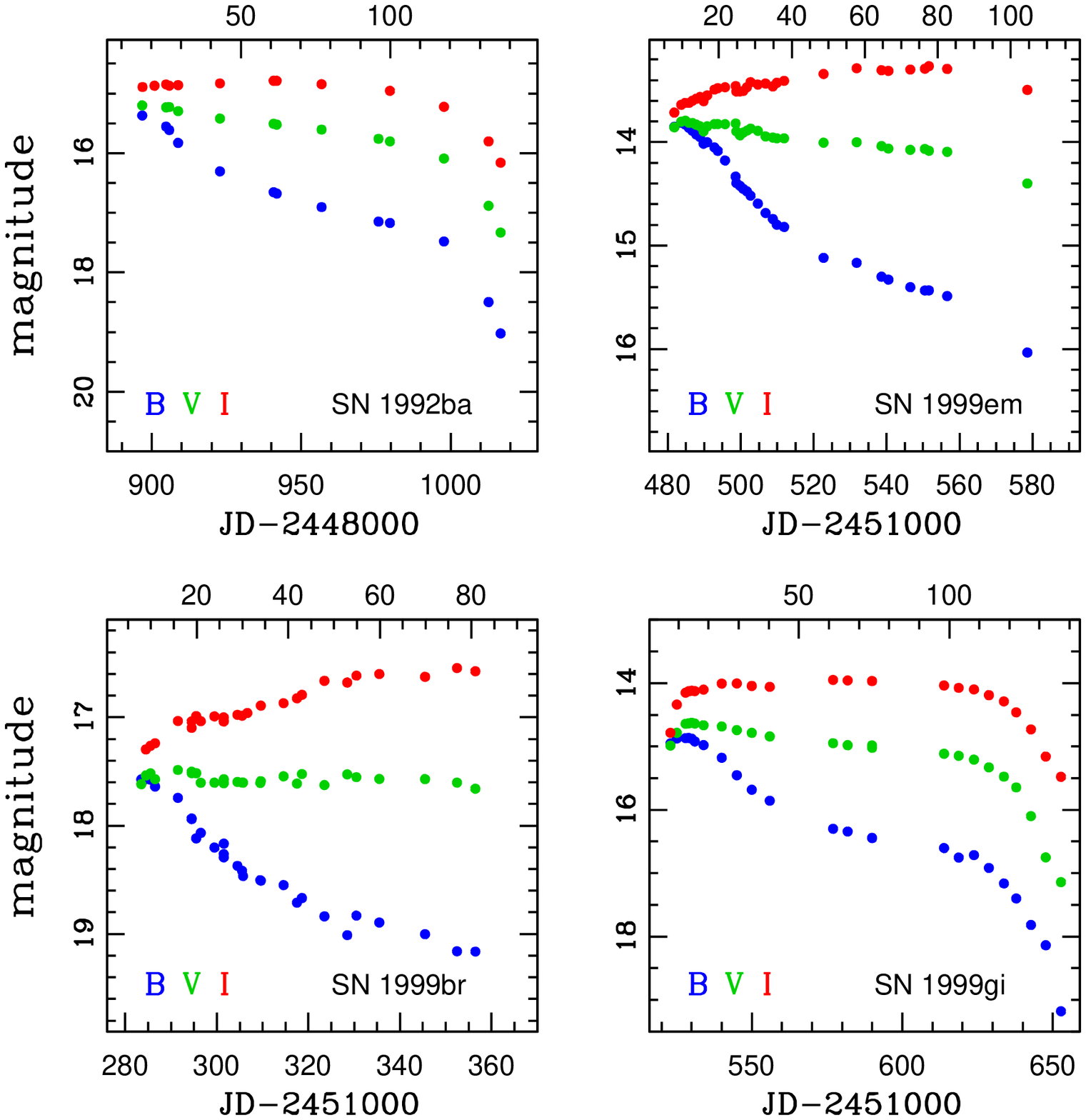}
\caption{Optical light curves of four SNe during the first 
$\sim\,120$ days of their evolution. The top axis of each panel gives
the phase in days since the EPM explosion time, derived using the {\rm
D05} models (see Table \ref{tab_VI_DES.EPM}).
~\label{fig_lightcurve1}}
\end{figure}

\begin{figure}
\plotone{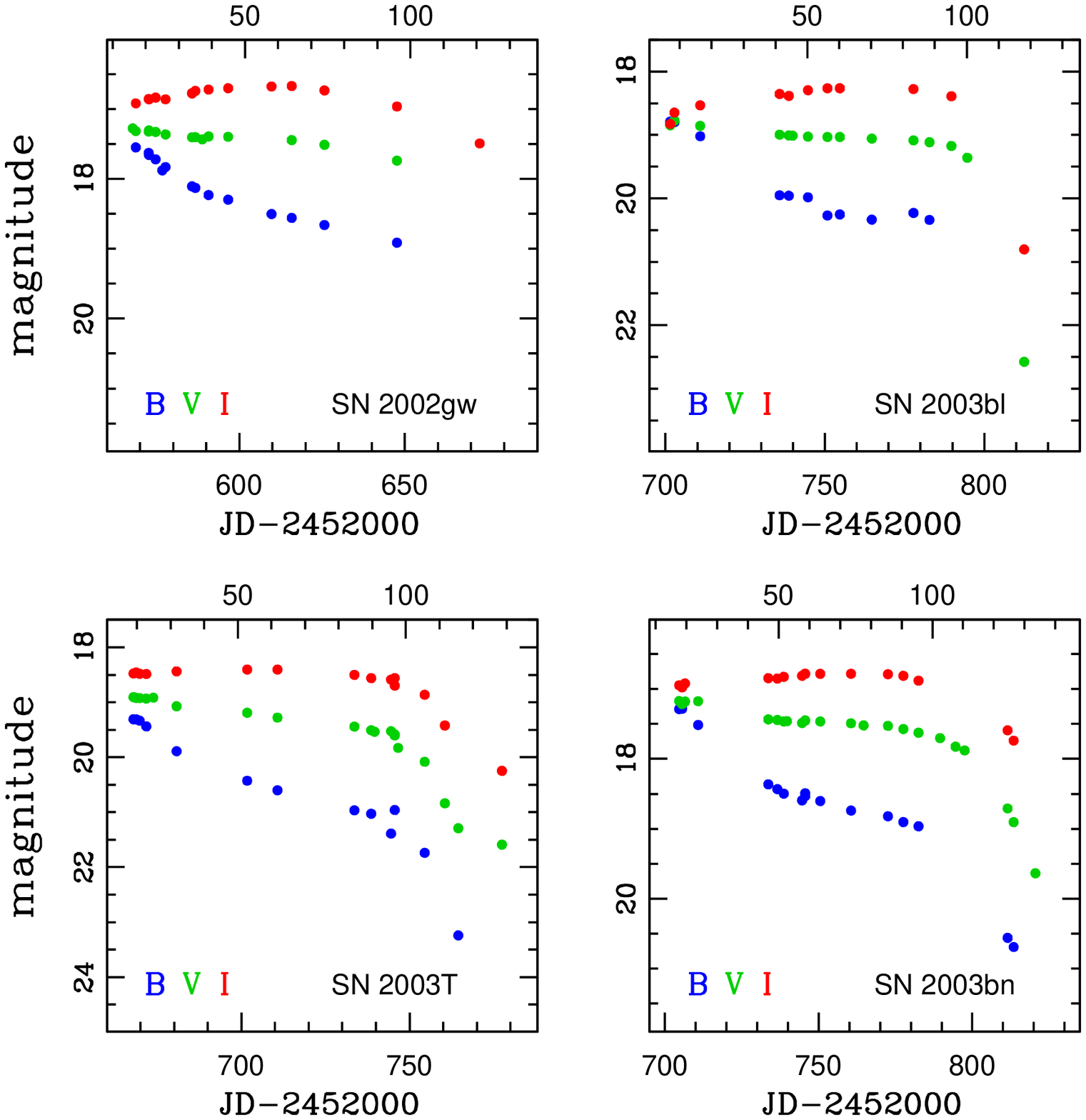}
\caption{Optical light curves of four SNe during the first 
$\sim\,120$ days of their evolution. The top axis of each panel gives
the phase in days since the EPM explosion time, derived using the {\rm
D05} models (see Table \ref{tab_VI_DES.EPM}).
~\label{fig_lightcurve2}}
\end{figure}

\begin{figure}
\plotone{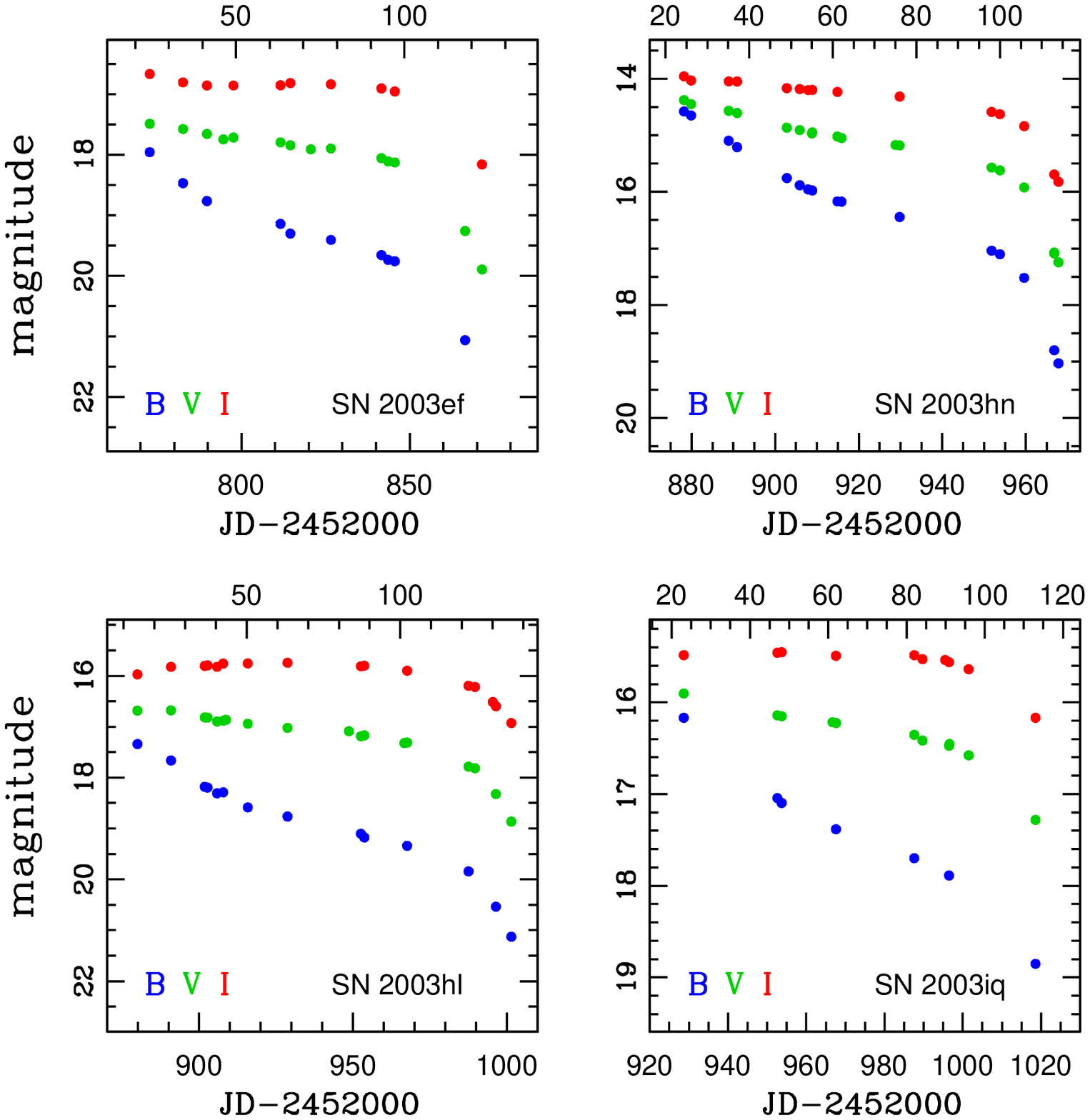}
\caption{Optical light curves of four SNe during the first 
$\sim\,120$ days of their evolution. The top axis of each panel gives
the phase in days since the EPM explosion time, derived using the {\rm
D05} models (see Table \ref{tab_VI_DES.EPM}).
~\label{fig_lightcurve3}}
\end{figure}

Direct images of SNe were obtained with telescopes from four
different observatories: the Cerro Tololo Inter-American Observatory
(CTIO), the Las Campanas Observatory (LCO), the European Southern
Observatory (ESO) in La Silla, and the Steward Observatory (S0).
Several telescopes and instruments were used to obtain the photometry,
which is listed in an electronic Table. In all cases CCD detectors
and standard Johnson-Kron-Cousins {\it U\,B\,V\,R\,I\,Z} filters
\citep{Jon66,Cou71} were employed. For a small subset of SNe
observations in the {\it JHK} filters were also obtained. The data
reduction was performed using IRAF\footnote{IRAF is distributed by the
National Optical Astronomy Observatory, which is operated by the
Association of Universities for Research in Astronomy, Inc., under
cooperative agreement with the National Science Foundation.} according
to the procedure described by \citet{Ham08}. The error in
the photometry ranges between 0.01-0.06 mag, with a typical value of
0.02 mag. 
\newline \indent
The optical light curves of all the SNe used in this work are shown in
Figures \ref{fig_lightcurve1}--\ref{fig_lightcurve3}, clearly
revealing the plateau nature of all these events. 

\subsection{Spectroscopy}

Low resolution ($R \sim 1000$) optical spectra (wavelength range
$\sim\,$3200--10000~\AA) were taken for each SN at various epochs using
telescopes and instruments from four different observatories. An
electronic Table lists all the telescopes and instruments used for the
spectroscopy. Most of the spectra were obtained with the slit along
the parallactic angle (Filippenko 1982). The wavelength calibration was
performed using comparison-lamp spectra taken at the position of
each SN. The flux calibration was done via observations of 
flux-standard stars \citep{Ham92,Ham94}. For more details on the
observational procedures see \citet{Ham08}.
\newline \indent
The spectra were shifted to the rest frame using the heliocentric
redshifts given in \mbox{Table \ref{tab_SN_list}} in order to measure
the SN ejecta velocities. In seven cases we were able to measure the
redshifts from narrow emission lines of H~II regions at the SN position
(see Table \ref{tab_SN_list}). Also, in one case (SN 1999em) we
adopted the value from \citet{Leo02b} which corresponds to the
redshift measured at the SN position. In four cases we were unable to
extract this information from our data, and we had to rely on redshifts
of the host-galaxy nuclei; this does not take into account the
rotation velocities of the host galaxies, which are typically $v\,\sim\,$
200 ${\rm km~ s^{-1}}$.

\subsection{Sample of Supernovae Used in this Work}

Fifty-one SNe~II were observed in the surveys described above. We cut this
sample according to the EPM requirements, which are (1) the optical
SN light curve ($V$ and $I$ bands) must show a nearly constant
luminosity during the photospheric phase, i.e, the SN must belong to
the SN~IIP class (see Figures
\ref{fig_lightcurve1}--\ref{fig_lightcurve3}); (2) the SN must have
early-time photometry; and (3) the SN must have at least three early
spectroscopic observations. The necessity for all of these requirements,
discussed in \S~3.6, reduced the sample to only 11 SNe. We also
added the SN IIP 1999gi to our sample, which has extensive photometry
and spectroscopy published by \citet{Leo02a}.

\section{The Expanding Photosphere Method}

\subsection{Basic Principles}

The EPM is a geometrical technique that relates an angular size and a
physical size of a SN, in order to derive its distance. Although the
angular radius $\theta$ of a SN cannot be resolved spatially with
current optical instrumentation, it can be derived assuming a
spherically symmetric expanding photosphere (a reasonable assumption
for SNe IIP at early times, as discussed by \citealt{L01}) that radiates
as a black body ``diluted" by a factor $\zeta^2$. Specifically,

\begin{equation}
\theta = \frac{R}{D} = \sqrt{\frac{(1+z)f_{\lambda}}{{\pi}{\zeta_{{\lambda}^{'}}^{2}}
B_{{\lambda}^{'} }(T)10^{-0.4[A(\lambda)+A^{'}({\lambda}^{'})]}}},
\end{equation} \newline
where $R$ is the photospheric radius, $D$ is the distance to the SN,
$f_{\lambda}$ is the observed flux density, $\lambda$ is the observed
wavelength, $B_{{\lambda}^{'}}$ is the Planck function in the SN rest
frame, $T$ is the color temperature, ${\lambda}^{'} = {\lambda}/(1+z)$
is the corresponding wavelength in the SN rest frame, $A(\lambda)$ is
the foreground dust extinction and $A^{'}({\lambda}^{'})$ is the
host-galaxy extinction. The factor $\zeta_{\lambda^{'}}$ (known as
``distance correction factor" or ``dilution factor") accounts for the
fact that a SN does not radiate as a perfect black body; there is flux
dilution caused by grey electron scattering which makes the
photosphere (defined as the region of total optical depth $\tau =
2/3$) form in a layer above the thermalization surface. Also, the
dilution factor accounts for line blanketing in the SN
atmosphere. Since electron scattering is the main source of continuum
opacity, the total opacity is essentially grey, and the photospheric
angular radius is independent of wavelength in the optical and
near-infrared \citep{E96}, which explains why $R$ and $\theta$ do not
carry a wavelength subscript.
\newline \indent
Because the gravitational binding energy ($U\,\sim\,10^{49}$ erg) of
a SN progenitor is far less than the expansion kinetic energy ($E\,
\sim\,10^{51}$ erg) of the ejecta, it is reasonable to assume free
expansion. This assumption is supported by hydrodynamical models
which show that the different layers of the ejecta reach $\sim\,$95\%
of their terminal velocities $\sim\,$1 day after the explosion. During
this brief period there is a transition from an acceleration phase due
to the SN explosion, to homologous expansion \citep{Utr07,Ber08}. Due
to the high expansion velocities ($\sim\,$10000 ${\rm km~s^{-1}}$), the
initial radius (typically $R_0 \sim\,10^{13}$ cm for a red
supergiant) can be neglected after $\sim\,$1 day from explosion; hence
after that period the physical radius of the SN can be approximated by

\begin{equation}
R \approx \frac{v(t-t_{0})}{1+z},
\end{equation} \newline 
where $v$ is the photospheric velocity and $t_{0}$ is the explosion
date. Combining (1) and (2) we obtain

\begin{equation}
\frac{\theta_{i}}{v_{i}} \approx \frac{(t_{i}-t_{0})}{(1+z)D},
\end{equation} \newline 
where $\theta_{i}$ and $v_{i}$ are the derived quantities measured at
time $t_{i}$, which are estimated following the steps explained in the
next sections. Equation 3 shows that the quantity $\theta/v$
increases linearly with time, so $D$ and $t_{0}$ can be derived from
at least two spectroscopic and photometric observations. More
observations allow us to check the internal consistency of the method.

\subsection{Dilution Factors}

\begin{figure}
\plotone{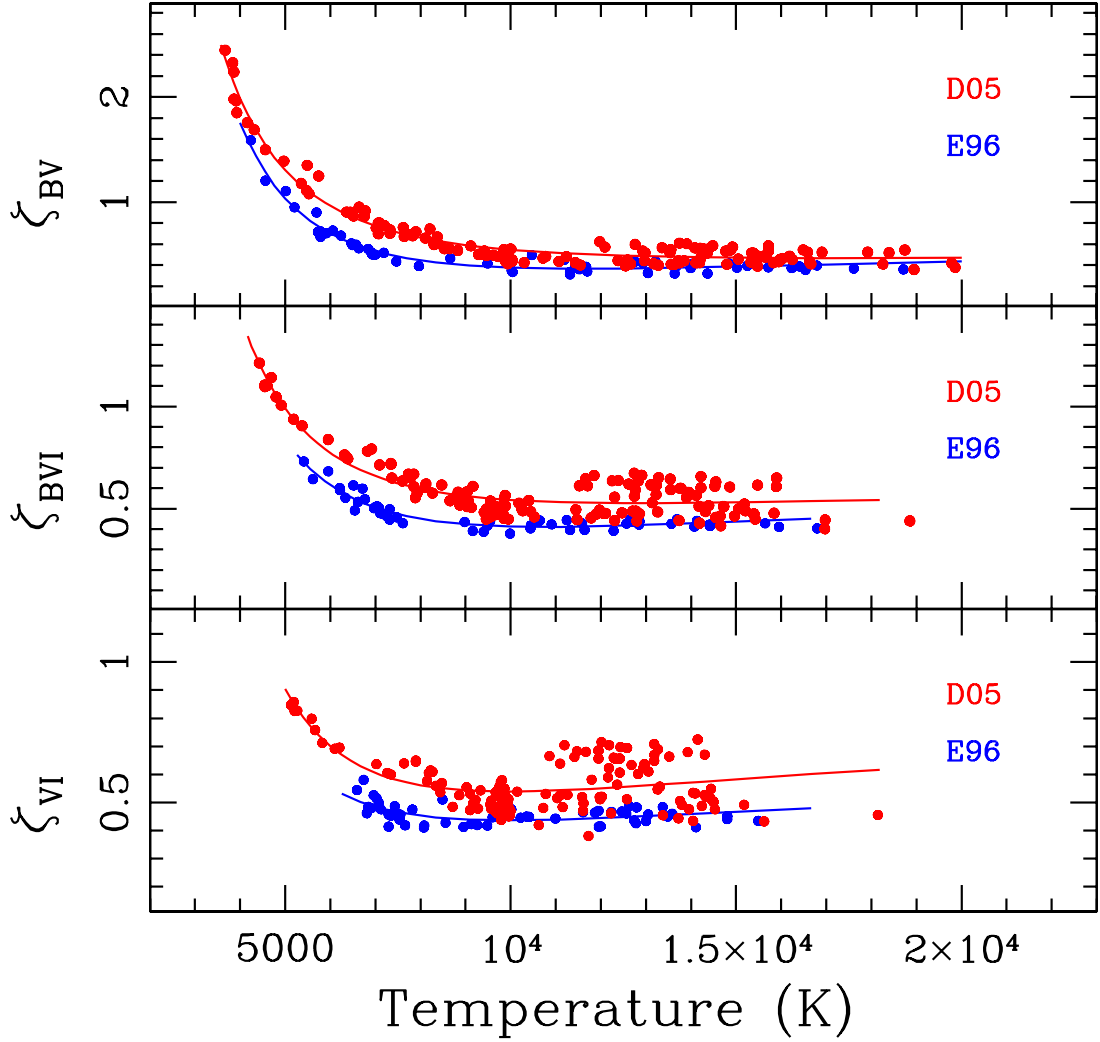}
\caption{Dilution factors $\zeta$ as a function of the color
temperature, computed at $z\,=\,0$ from the {\rm E96} (blue dots) and
{\rm D05} (red dots) atmosphere models for three different filter
subsets ($\{BV\},\,\{BVI\},\,\{VI\}$).  The blue (red) line corresponds to
the polynomial fit performed to the {\rm E96} ({\rm D05}) dilution
factors.
~\label{fig_zeta}}
\end{figure}

The dilution factors correspond to the ratio of the luminosity of a SN
atmosphere model ($L_{\lambda^{'}}$) and the corresponding black-body
luminosity,

\begin{equation}
\zeta_{\lambda^{'}}^2 = \frac{L_{\lambda^{'}}}{{\pi}B_{\lambda^{'}}(T)4{\pi}R^2}.
\end{equation} \newline
In practice, the dilution factors must be derived for the same filter
subsets employed to determine the color temperature ($T$) of a SN. In
this work we focus on three different optical filter subsets
($\{BV\}$, $\{BVI\}$ and $\{VI\}$), and we used two SN atmosphere
models, those by \citet{E96} ({\rm E96} hereafter) and
\citet{D05b} ({\rm D05} hereafter), to compute the dilution
factors. See also \citet{D05a} for more details of the input
parameters of the {\rm D05} models. Because the color temperature of
each SN was determined from colors measured in the observer's rest
frame, both the atmosphere models and the black-body function must be
redshifted; thus, the dilution factors must be computed for the
specific redshift of each SN.
\newline \indent
We computed {\it B,V,I} synthetic magnitudes using 58 spectra from
{\rm E96} atmosphere models and 138 spectra from {\rm D05} atmosphere
models. For each filter subset $S$ (that is, $S = \{BV\},\, \{BVI\},\,
\{VI\}$), we fit black-body functions in the SN rest frame
$B_{\lambda^{'}}(T_s)$, and solved for $T_s$ and $\zeta_{S,z}$ by
minimizing the quantity

\begin{equation}
\epsilon=\sum_{{\overline{\lambda}} \in S}{[M_{\overline{\lambda}}} + 5\log(\frac{R}{10~{\rm pc}}) + 
5\log(\zeta_{S,z}) - b_{\overline{\lambda}}(T_s,z)]^2.
\end{equation} \newline
Here $R$ is the photospheric radius, $M_{\overline{\lambda}}$ is the
redshifted synthetic absolute magnitude of the atmosphere model for a
band with central wavelength ${\overline{\lambda}}$, and
$b_{\overline{\lambda}}(T_s,z)$ is the synthetic magnitude of
${\pi}B_{\lambda^{'}}
(T_s)10^{-0.4[A(\lambda)+A^{'}(\lambda^{'})]}/(1+z)$, given by

\begin{equation}
b_{\overline{\lambda}} = -2.5\log_{10}\int\frac{{{\pi{\lambda}B_{\lambda^{'}}(T_s)
10^{-0.4[A(\lambda)+A^{'}({\lambda}^{'})]}}}}{hc(1+z)}S(\lambda)d\lambda + ZP,
\end{equation} \newline
where $S(\lambda)$ is the filter transmision function and $ZP$ is the
zero point of the photometric system \citep{Ham01}. The constants $h$
and $c$ are the Planck constant and the speed of light,
respectively. Clearly, the dilution factors depend on the specific
redshift of the SN and on the filter subset used to obtain
temperature of the models. Figure \ref{fig_zeta} shows the resulting
dilution factors versus temperature at $z\,=\,0$. We performed
polynomial fits to $\zeta(T_s)$ of the form

\begin{equation}
\zeta(T_s) = \sum_{j=0}^{2} b_{s,j}\left( \frac{10^{4}~{\rm K}}{T_s} \right )^{j}.
\end{equation} \newline
Table \ref{tab_zeta_coef} lists the $b_{s,j}$ coefficients at
$z\,=\,0$ for three filter subsets and both atmosphere models (E96 and
D05). The corresponding polynomial fits are shown in \mbox{Figure
\ref{fig_zeta}}.
\newline \indent
The {\rm D05} dilution factors are quite insensitive to the color
temperature above $\sim\,$9000 K, and lie around 0.5, while at lower
temperatures they increase sharply with decreasing temperature,
reaching a value over unity below $\sim\,$5000 K. The {\rm E96}
dilution factors present the same pattern, but they are systematically
lower than the {\rm D05} dilution factors by \mbox{$\sim\,$${\rm
15\,\%}$.} The origin of these differences is unclear. \citet{D05a}
discuss that the discrepancy might be related to the different
approach used to handle relativistic terms. Also, {\rm D05} solved
the non-LTE (local thermodynamic equilibrium) (non-LTE) problem for
all the species, and employed a very complex atom model. {\rm E96}, on
the other hand, solved the non-LTE problem for a few species, while
for the rest of the metals the excitation and ionization were assumed
to be given by the Saha-Boltzmann equation, and the opacity was taken
as pure scattering. Another important difference between the {\rm
E96} and {\rm D05} dilution factors is the dependence on the
parameters involved in the atmosphere modelling. While the {\rm E96}
dilution factors show little sensivity to a broad range of phyical
parameters other than temperature, the {\rm D05} models show a larger
dispersion at a given color temperature. However, this is also due to
{\rm D05} models covering a larger range of radii, density profiles
(${\rm \rho\, \propto\, r^{-n}}$) and temperature than {\rm E96}. 
On average, the {\rm E96} models lead to a dispersion of $\sigma\,\sim\,    
0.03$ in $\zeta$, while the {\rm D05} models yield $\sigma\,\sim\,0.07$.

\subsection{Angular Radii}

An apparent angular radius ($\theta\zeta_{s}$) and a color temperature
($T_s$) of the SN can be obtained by fitting a Planck function
$B_{{\lambda^{'}}}(T_{s})$ to the observed broad-band magnitudes (see
eq. 1). Here $S$ is the filter subset combination, i.e., $S = \{BV\},\,
\{BVI\},\, \{VI\}$. Since we have two unknowns
($\theta\zeta_{s}$,$T_{s}$), the subsets must contain at least two
filters. In order to derive these parameters, we used a least-squares
technique at each spectroscopic observation epoch (see \S 3.6) by
minimizing the quantity

\begin{equation}
\chi^{2} = \sum_{s}\frac{[m_{\overline{\lambda}} + 5\,{\rm log}(\theta\zeta_{s,z})-
b_{\overline{\lambda}}(T_s,z)]^2}{\sigma_m^2}.
\end{equation} \newline
Here, $m_{\overline{\lambda}}$ is the apparent magnitude in the filter
with central wavelength $\overline{\lambda}$ (i.e.,
$m_{\overline{\lambda}}\in\{B,V,I\}$), $\sigma_{m}$ is the photometric
error in the magnitude $m_{\overline {\lambda}}$, and
$b_{\overline{\lambda}}$ is defined in eq. 6. Because $\zeta_s$ is
mainly a function of the color temperature (Fig. \ref{fig_zeta}), it
is possible to use $T_s$ to solve for $\zeta_{s}$ and determine the
true angular radius $\theta$ from $\theta\zeta_{s}$.

\subsection{Physical Radii}

\begin{figure}
\plotone{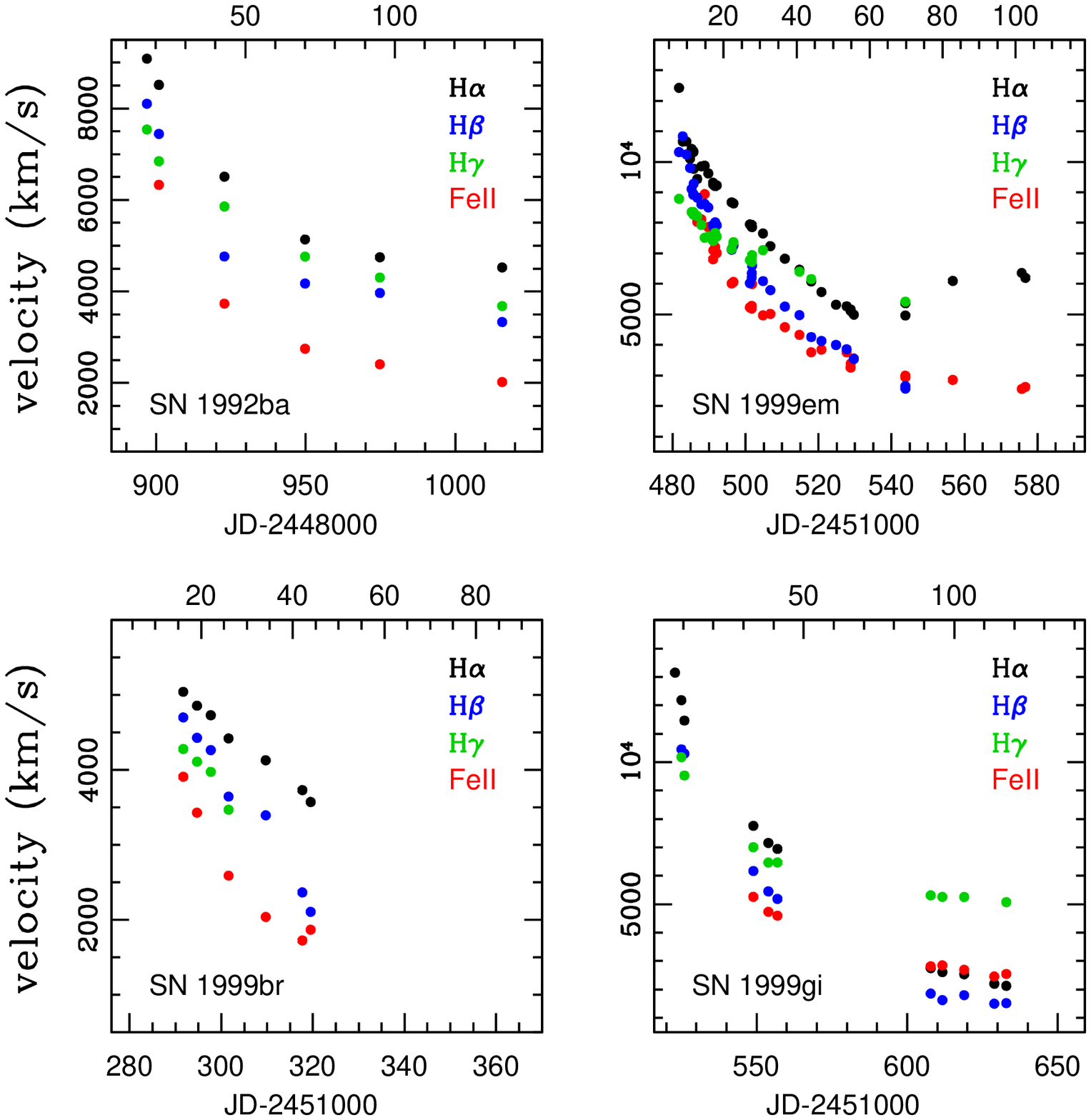}
\caption{Line velocity evolution determined from the P Cygni
absorption minima of four different features during ${\rm \sim \,
100}$ days after discovery. The top axis of each panel gives the phase
in days since the EPM explosion time derived using the {\rm D05}
models (see Table \ref{tab_VI_DES.EPM}).
~\label{fig_SNe_velocities1}}
\end{figure}

\begin{figure}
\plotone{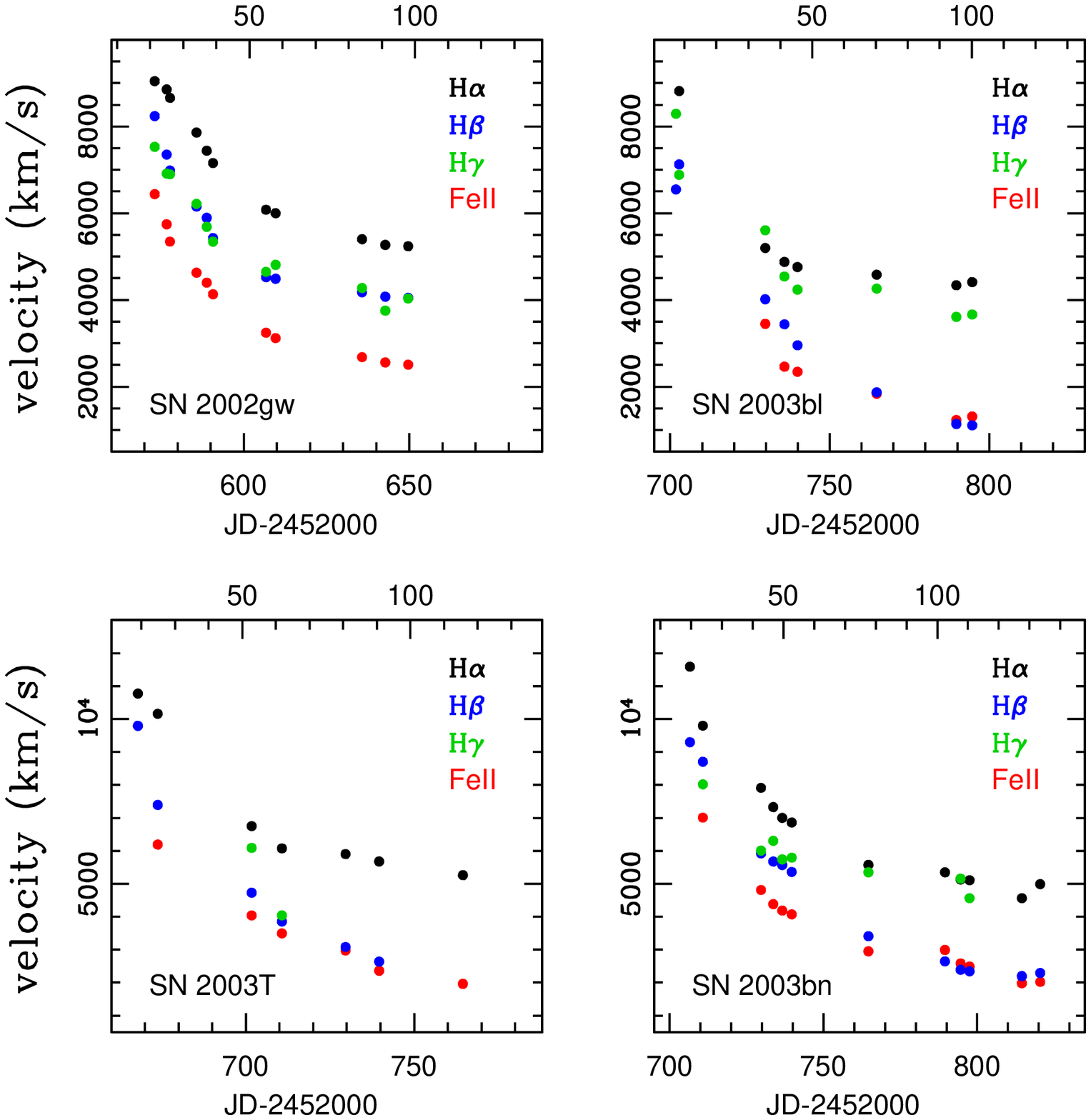}
\caption{Line velocity evolution determined from the P Cygni
absorption minima of four different features during ${\rm \sim \,
100}$ days after discovery. The top axis of each panel gives the phase
in days since the EPM explosion time derived using the {\rm D05}
models (see Table \ref{tab_VI_DES.EPM}).
~\label{fig_SNe_velocities2}}
\end{figure}

\begin{figure}
\plotone{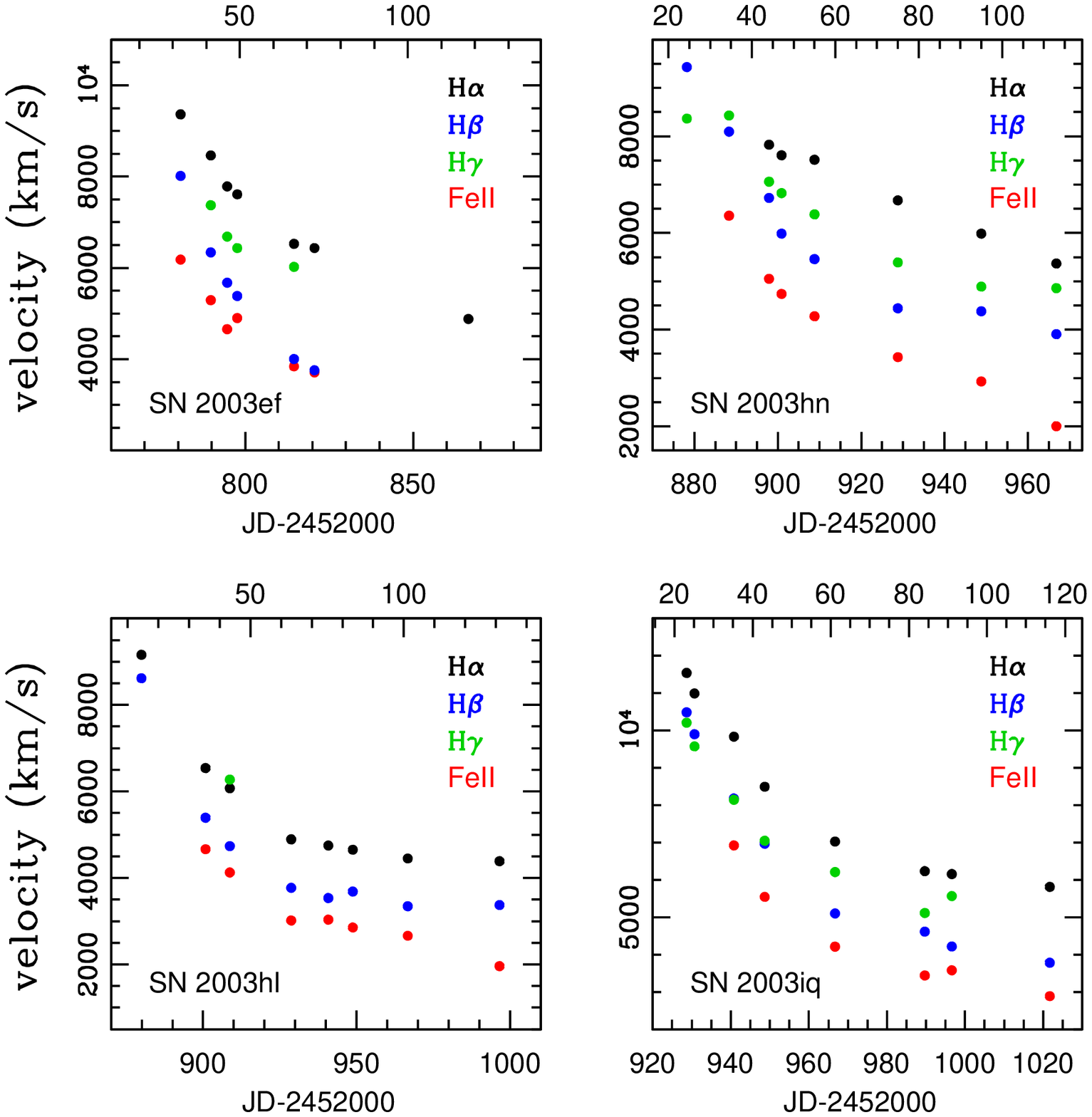}
\caption{Line velocity evolution determined from the P Cygni
absorption minima of four different features during ${\rm \sim \,
100}$ days after discovery. The top axis of each panel gives the phase
in days since the EPM explosion time derived using the {\rm D05}
models (see Table \ref{tab_VI_DES.EPM}).
~\label{fig_SNe_velocities3}}
\end{figure}

\begin{figure}
\plotone{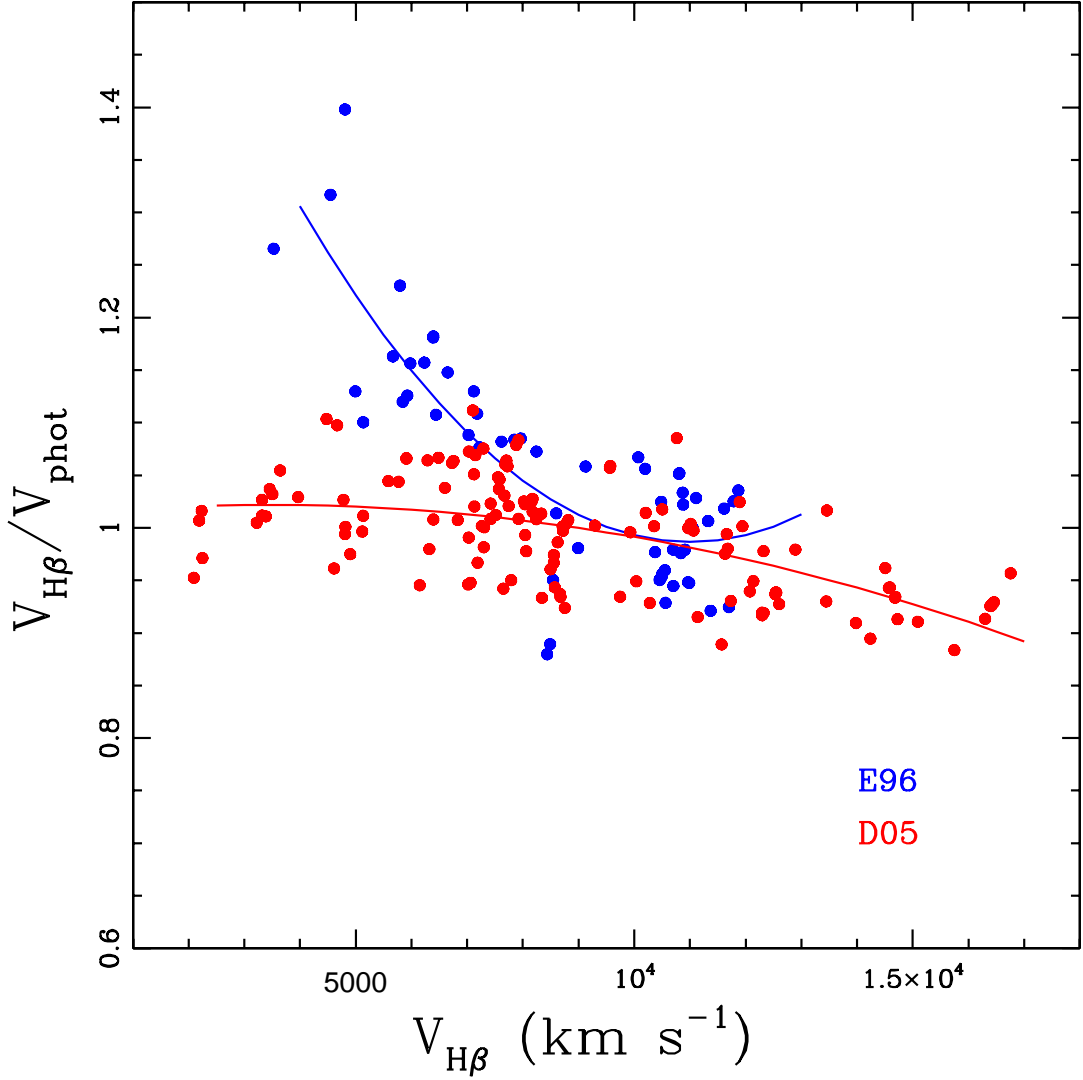}
\caption{Ratio of the H$\beta$ velocity to the photospheric velocity
versus the H$\beta$ velocity of the individual SN models. The blue
dots correspond to {\rm E96} models and the red dots to {\rm D05}
models. The blue (red) line corresponds to the polynomial fit
performed to the {\rm E96} ({\rm D05}) photospheric velocity
conversion.
~\label{fig_models_VHbeta_Vphot}}
\end{figure}

Once $\theta$ is determined, the next step is to measure the
photospheric velocity (see eq. 3). The photospheric velocity of the SN
at a given epoch can be obtained from the absorption lines in the
spectra. We measured velocities\footnote{We employed the non-relativistic 
formulae to derive the expansion velocities from the doppler shift      
of the absorption lines. This approach is reasonable because the highest 
velocity used in this work is $\sim$0.035c, for which the difference 
in the velocity using the relativistic and non-relativistic formulae
is less than 2\%. However, typical velocities employed in the EPM analysis
are of 0.02c, for which the difference is $\sim$ 1\%.} from the minima of 
H$\alpha$, H$\beta$, H$\gamma$, and Fe {\sc ii} $\lambda\,5169$ lines, for
all 12 SNe. The tables that list the spectroscopic velocities are
available in electronic format. Figures
\ref{fig_SNe_velocities1}-\ref{fig_SNe_velocities3} show the temporal
evolution of the spectral line velocities.
\newline \indent
To date the photospheric velocities have been estimated using weak
spectral absorption features such as Fe {\sc ii}
$\lambda \lambda5169$, 5018, 4924, and Sc {\sc ii} $\lambda 4670$
\citep{SKE92,Leo02b}. The physical assumption is that these lines are
weak and formed near the photosphere of the SN. However,
there are two problems with this approach: (1) at early times the
spectra are dominated by Balmer lines and the weak lines are absent,
and (2) the synthetic spectra show that even the weak lines do not
necessarily yield true photospheric velocities \citep{D06}. One way to
circumvent these problems is to use the Balmer lines which are present
in the spectra over most of the evolution of the SN. Although the
Balmer lines are optically thicker than the \mbox{Fe {\sc ii}} lines,
\citet{D06} argue that, contrary to what is usually believed,
optically thick lines do not necessarily overestimate the photospheric
velocity, and the offset from the photospheric velocity can be
measured from the synthetic spectra. In this work we decided to use
the minimum of the H$\beta$ absorption line to derive the photospheric
velocity because this line is present during the entire plateau phase,
it can be easily identified, and it does
not present any blend, at least in the first $\sim\,$50 days
after explosion.
\newline \indent
To convert from observed H$\beta$ spectroscopic velocities to true
photospheric velocities we used the synthetic spectra from {\rm E96}
and {\rm D05}. Figure \ref{fig_models_VHbeta_Vphot} shows (in red)
the ratio of H$\beta$ velocity and the photospheric velocity, as a
function of H$\beta$ velocity for all of the {\rm D05} models. Note that
the {\rm D05} models predict that the H$\beta$ line forms quite close
to the photosphere at all epochs (for all values of $v_{H\beta}$).
Also plotted in Figure \ref{fig_models_VHbeta_Vphot} (in blue) are the
{\rm E96} models which confirm that the H$\beta$ forms close to the
photosphere at early epochs, when $v_{H\beta}$ is high. However,
at later epochs (lower $v_{H\beta}$) {\rm E96} predict that H$\beta$
forms in outer layers (higher velocities) than {\rm D05}. It is also
important to note that the {\rm E96} models cover a shorter range in
velocity (${\rm \sim\,}$4500--12000 ${\rm km~ s^{-1}}$) than the {\rm
D05} models ($\sim\,$2000--17000 ${\rm km~ s^{-1}}$), which
restricts the EPM analysis using the {\rm E96} models.
\newline \indent
To derive the ratio between the H$\beta$ and the photospheric velocity
we used a polynomial fit of the form

\begin{equation}
\frac{v_{H\beta}}{v_{phot}} = \sum_{j=0}^{2} a_{j}(v_{H\beta})^{j}, 
\end{equation} 
(see Figure \ref{fig_models_VHbeta_Vphot}). The $a_{j}$ coefficients
are listed in Table \ref{tab_ratio_coef}. The {\rm E96} models lead to
a dispersion of ${\rm \sigma\, =\, 0.06}$ and the {\rm D05} models to
${\rm \sigma\, =\, 0.04}$. The photospheric velocity $v_{i}$ can be
obtained from a measurement of $v_{H\beta}$:
 
\begin{equation}
v_{i} = \frac{v_{H\beta}}{\displaystyle{\sum_{j=0}^{2} a_{j}(v_{H\beta})^{j}}}.
\end{equation}

\begin{figure}
\plotone{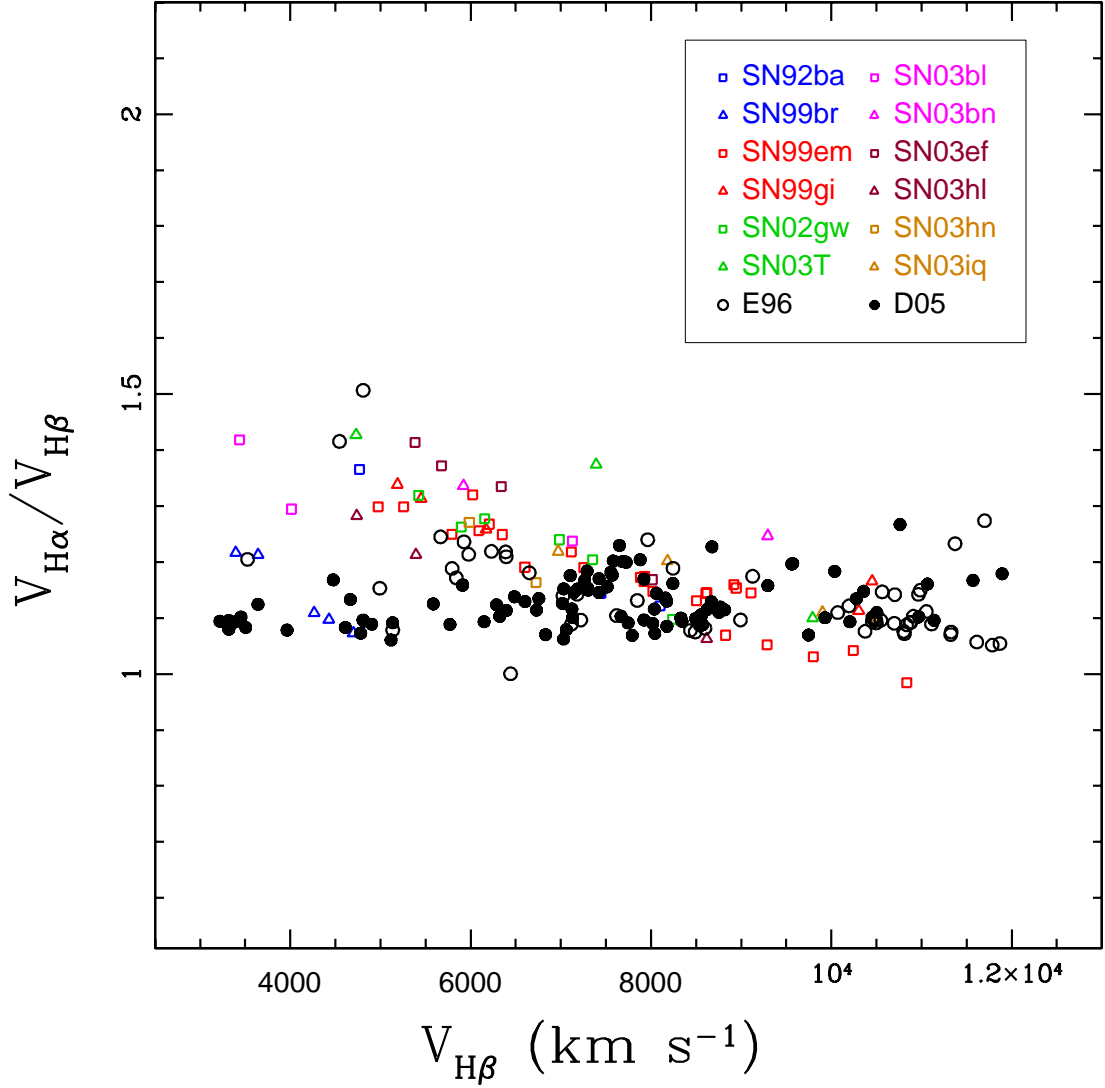}
\caption{Ratio of the H$\alpha$ velocity to the H$\beta$ velocity as 
a function of the H$\beta$ velocity. The triangles and the squares
represent velocities measured from the spectra of our SN sample. The
open and filled black circles correspond to the velocity ratio
measured from the synthetic spectra of {\rm E96} and {\rm D05},
respectively.
~\label{fig_obs_vel.ratio}}
\end{figure}

In order to examine which of the adopted photospheric velocity
conversions was closer to reality, we compared the ratio between the
H$\alpha$ and H$\beta$ velocities measured from the observed spectra
of our sample of SNe and from the synthetic spectra of the {\rm E96}
and {\rm D05} models. Figure \ref{fig_obs_vel.ratio} shows the
H$\alpha/H\beta$ velocity ratio as a function of the H$\beta$
velocity. It can be seen that, while there is good agreement between
theory and observations at high H$\beta$ velocities ($\sim\,$7000--10500
${\rm km~ s^{-1}}$), the {\rm D05} models underestimate the
H$\alpha$ velocities (or overestimate the H$\beta$ velocities) at
lower expansion velocities. This could be due to time-dependence
effects that are not include in the {\rm D05} models that assume
steady-state \citep{D08b}. On the other hand, the H$\alpha$/H$\beta$
velocity ratio predicted by the {\rm E96} models is in good agreement
with the observations at all H$\beta$ velocities, although there are
few models below  $\sim\,$6000 ${\rm km~ s^{-1}}$ to draw strong conclusions.
This suggests that {\rm E96} predict more realistic line profiles in the SN 
ejecta than {\rm D05} and therefore should provide a better photospheric 
velocity conversion.

\subsection{Extinction}      

\begin{figure}
\plotone{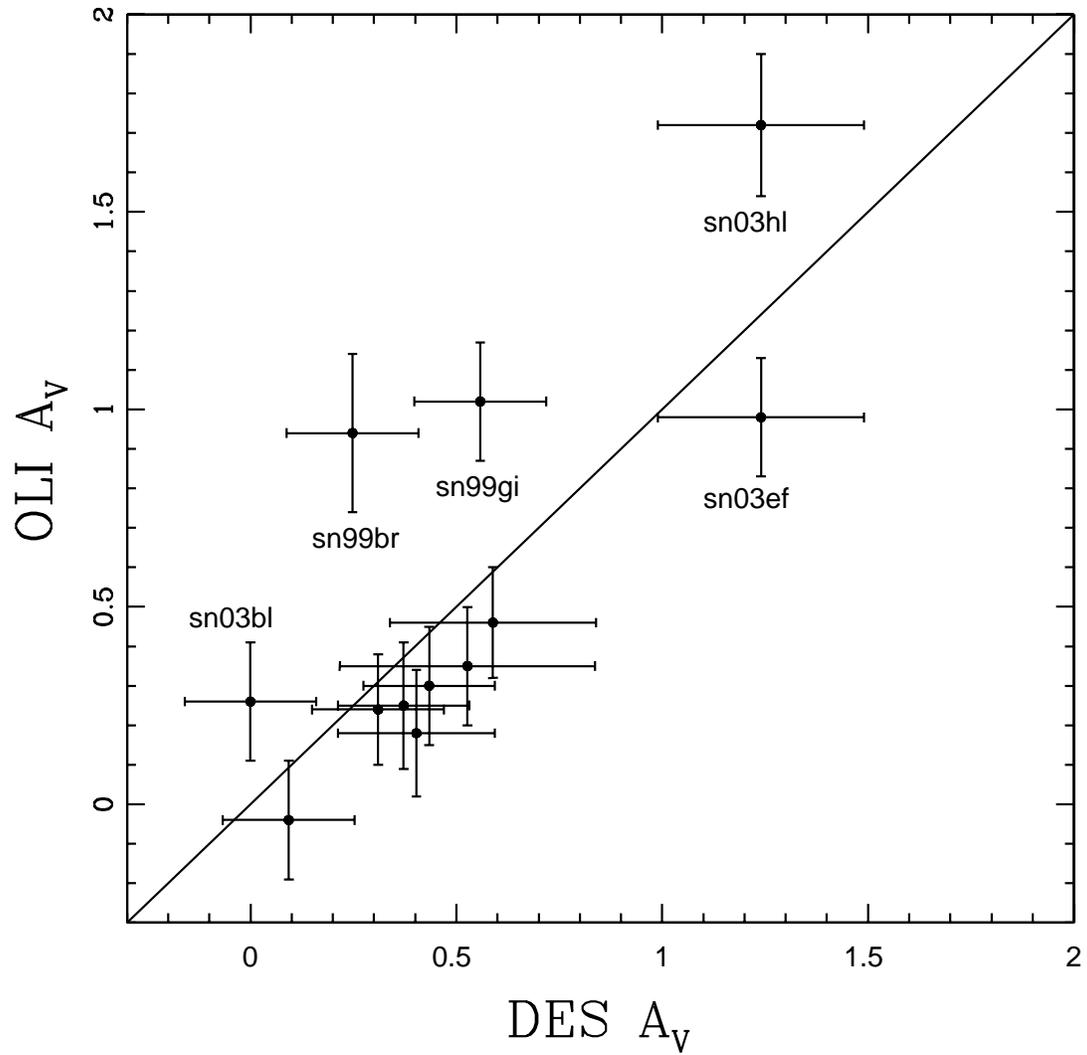}
\caption{Comparison between the {\rm DES} and {\rm OLI} reddening
methods for the 12 SNe.  The straight line has a slope of unity. The
more deviant SNe are labeled.
~\label{fig_DES_OLI_reddening}}
\end{figure}

To estimate the amount of Galactic foreground extinction we used the
IR dust maps of \citet{Sch98}. Table \ref{tab_reddening} summarizes
the foreground extinction adopted. In this work we used two different
methods for the determination of host-galaxy reddenings of our SN
sample: a spectroscopic method ({\rm DES} hereafter), and a method
based on the color evolution of the SNe ({\rm OLI} hereafter). The
former consists in fitting different model spectra to the early-time
spectra of a SN. The two fitting parameters are the amount of
reddening and the photospheric temperature \citep{D06,D08a}. The
color-based technique was developed by \citet{OLI08} and is based on
the assumption that the color at the end of the plateau phase is the
same for all SNe IIP.
In both cases we adopted the \citet{Car89} extinction law (with
$R_{V}=3.1$).
\newline \indent
Table \ref{tab_reddening} lists the host galaxy visual extinction
values ${\rm A_V}$ obtained from both methods. Also, Figure
\ref{fig_DES_OLI_reddening} shows the {\rm OLI} versus {\rm DES}
visual extinctions. As can be seen, there are no systematic
differences between the models. However, there are individual
differences, especially in five SNe, whose names are explicitly marked
in the plot.

\subsection{Implementation of EPM}

\begin{figure}
\plotone{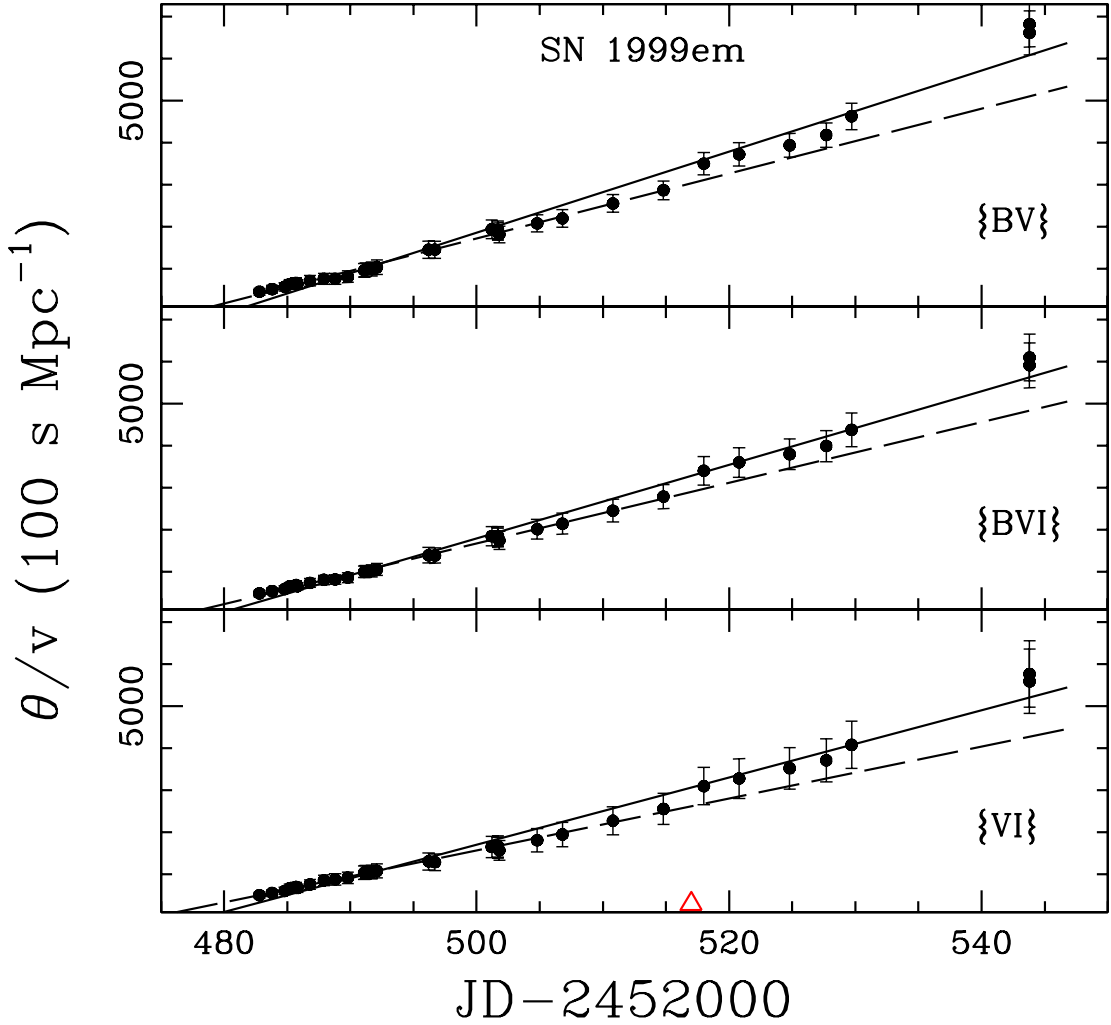}
\caption{The ratio $\theta/v$ as a function of time for SN 1999em
using the $\{BV\}$, $\{BVI\}$, and $\{VI\}$ filter subsets and the {\rm
D05} models. The solid (dashed) lines correspond to unweighted
least-squares fits to the derived EPM quantities using ${\rm \sim\,
70}$ (40) days after explosion.  The red triangle in the bottom panel
shows day ${\rm \sim\, 40}$ after explosion.
~\label{fig_full_SN99em.EPM}}
\end{figure}

The EPM method is only valid in the optically thick phase of a H-rich
expanding atmosphere. Observationally, this period corresponds to the
plateau phase of Type II SNe and thus justifies our first selection
criterion in $\S$ 2.3.
\newline \indent
 The EPM requires at least two simultaneous photometric and
spectroscopic observations (see eq. 3), but we recommend the use of at
least three points in order to obtain an internal check. The
photometry is used to determine the angular size of the SN and the
spectroscopy is used to measure the expansion velocities of the SN.
The requirement of simultaneous photometric and spectroscopic
observations is usually not accomplished; the photometry and the
spectroscopy of a SN are taken at different epochs. To overcome this
problem, it is necessary to interpolate the photometry or the
velocities measured from the spectra. In this work we decided to
interpolate the photometry for two reasons: (1) the number of
photometric observations in our sample of SNe is far greater than the
number of spectroscopic observations, and (2) the optical apparent
magnitude of SNe~II-P is nearly constant during the plateau phase,
making the photometry interpolation more reliable than the velocity
interpolation, which has a steeper dependence with time. To
interpolate a magnitude at the epoch of a given spectroscopic
observation, we use a quadratic polynomial fit, using four 
photometric observations around the spectroscopic
date.
\newline \indent
In this study, we restricted the EPM analysis to the first 
$\sim\,$45--50 days after explosion because there are clear departures
from linearity in the $\theta/v$ versus $t$ plots after this
date. \citet{D05b} show that the {\rm D05} models are poor at late
time and therefore should not be used for such epochs, supporting this
restriction. In Figure \ref{fig_full_SN99em.EPM} we plot the EPM
solutions for SN 1999em (because it has extensive photometric and
spectroscopic observations during the plateau phase) using 
the $\{BV\},\, \{BVI\}$, and $\{VI\}$ filter subsets and the {\rm
D05} models. The solid line corresponds to the least-squares fit to
the derived EPM quantities using the first ${\rm \sim\,70}$ days
after explosion, while the dashed line correspond to the least-squares
fit using only the first ${\rm \sim\,40}$ days after explosion. As
can be noted, after ${\rm \sim\,40}$ days from explosion (marked with
a red triangle) there is departure from the linear $\theta/v$ versus
$t$ relation in all three cases. This justifies our second and third
selection criteria in \mbox{$\S$ 2.3}. However, this restriction
severely lowers the number of SNe of our sample to which we can apply
the EPM. Out of the initial 51 SNe of the \citet{Ham08} sample, only
11 objects fulfill the requirement of having a plateau behavior and
having early-time photometry and spectroscopy for the EPM analysis.

\subsubsection{EPM Analysis of Individual SNe}

\begin{figure}
\plotone{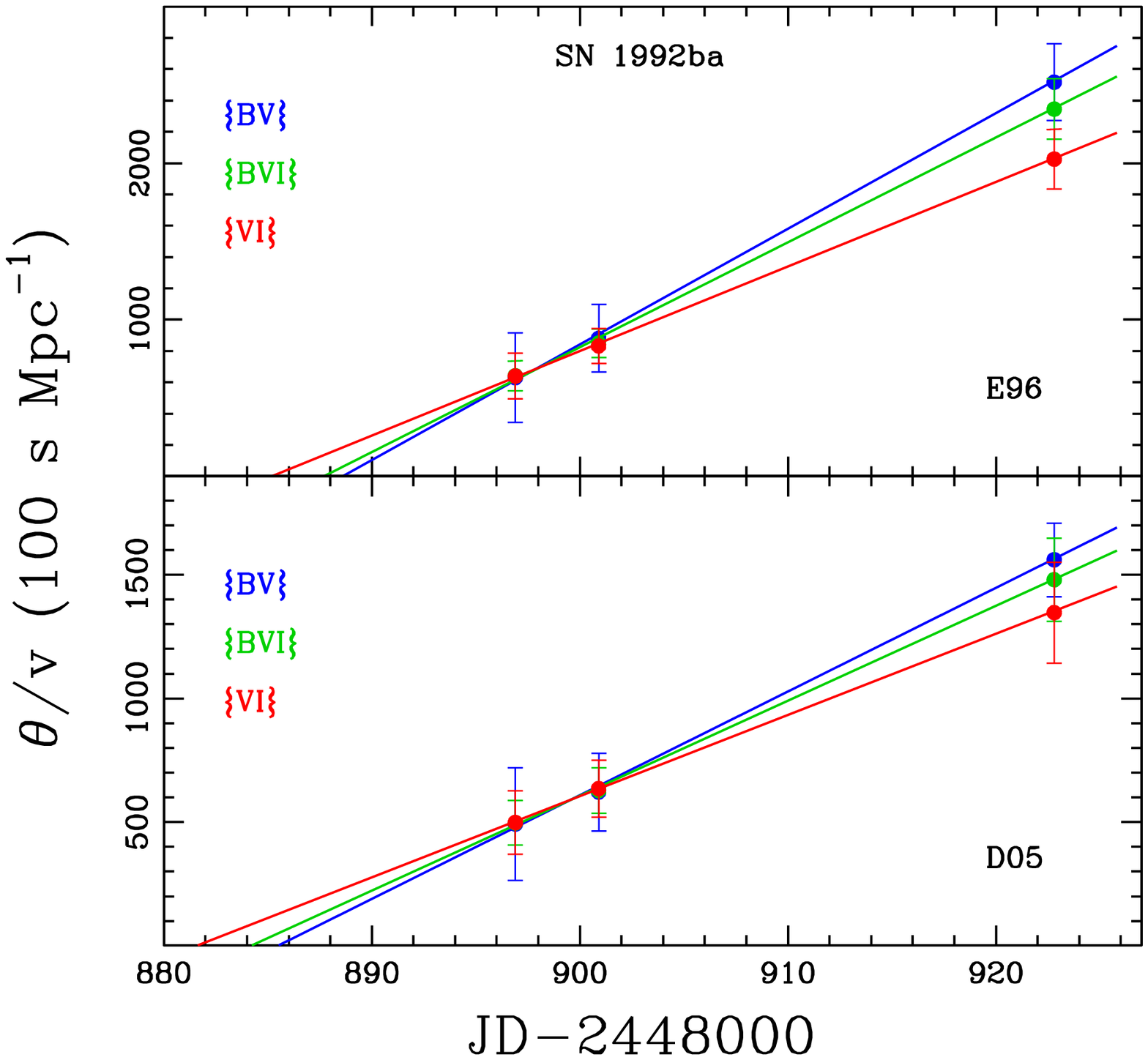}
\caption{The ratio $\theta/v$ as a function of time for SN 1992ba
using the $\{BV\}$, $\{BVI\}$, and $\{VI\}$ filter subsets. The ridge
lines correspond to unweighted least-squares fits to the derived EPM
quantities. The upper and lower panels show the results using {\rm
E96} and {\rm D05} dilution factors, respectively. In all cases we
employ the {\rm DES} reddening.
~\label{fig_SN92ba_EPM}}
\end{figure}

\begin{figure}
\plotone{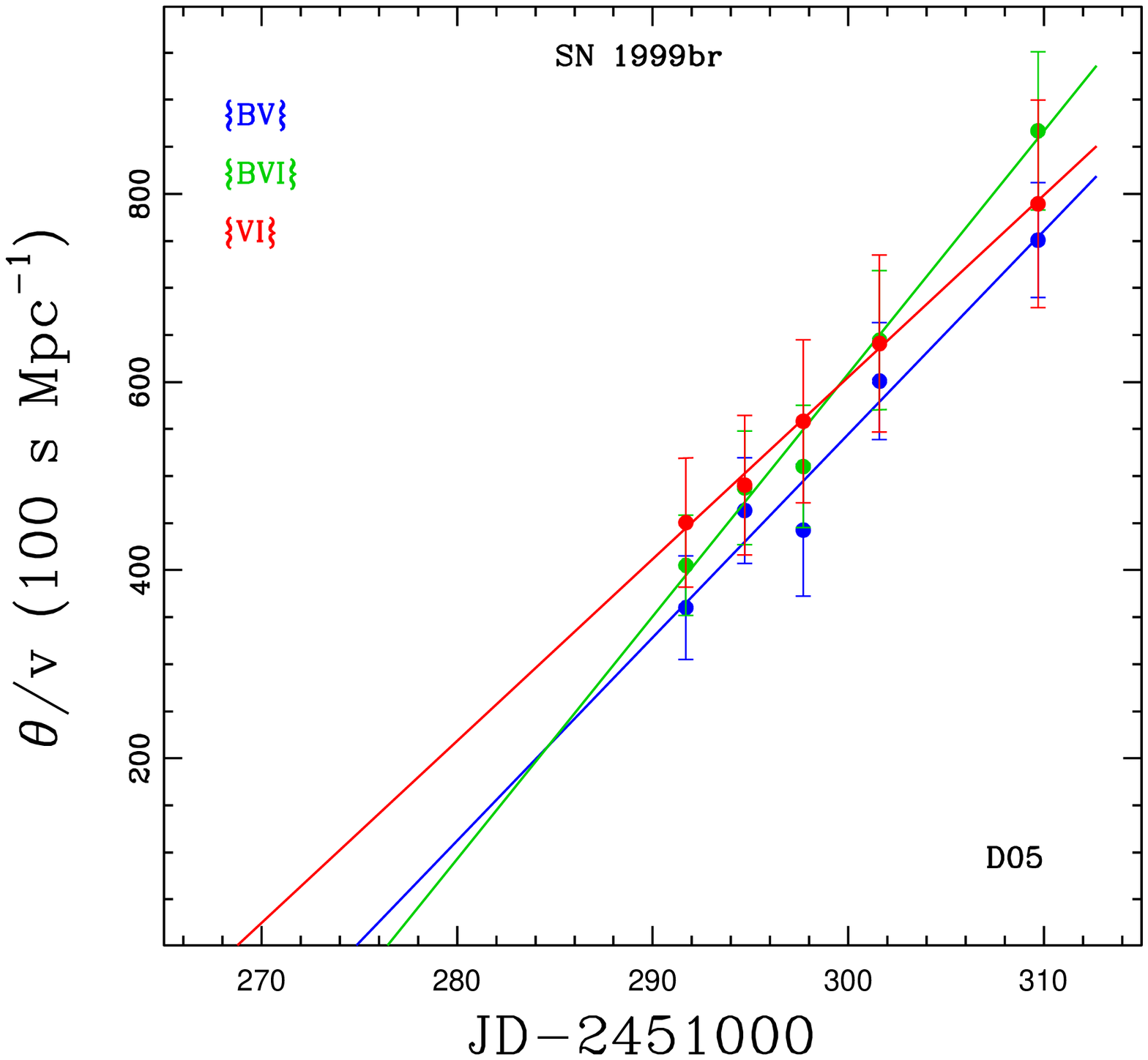}
\caption{The ratio $\theta/v$ as a function of time for SN 1999br
using the $\{BV\}$, $\{BVI\}$, and $\{VI\}$ filter subsets and the {\rm
D05} models. The ridge lines correspond to unweighted least-squares
fits to the derived EPM quantities. In all cases we employ the {\rm
DES} reddening.
~\label{fig_SN99br_EPM}}
\end{figure}

\begin{figure}
\plotone{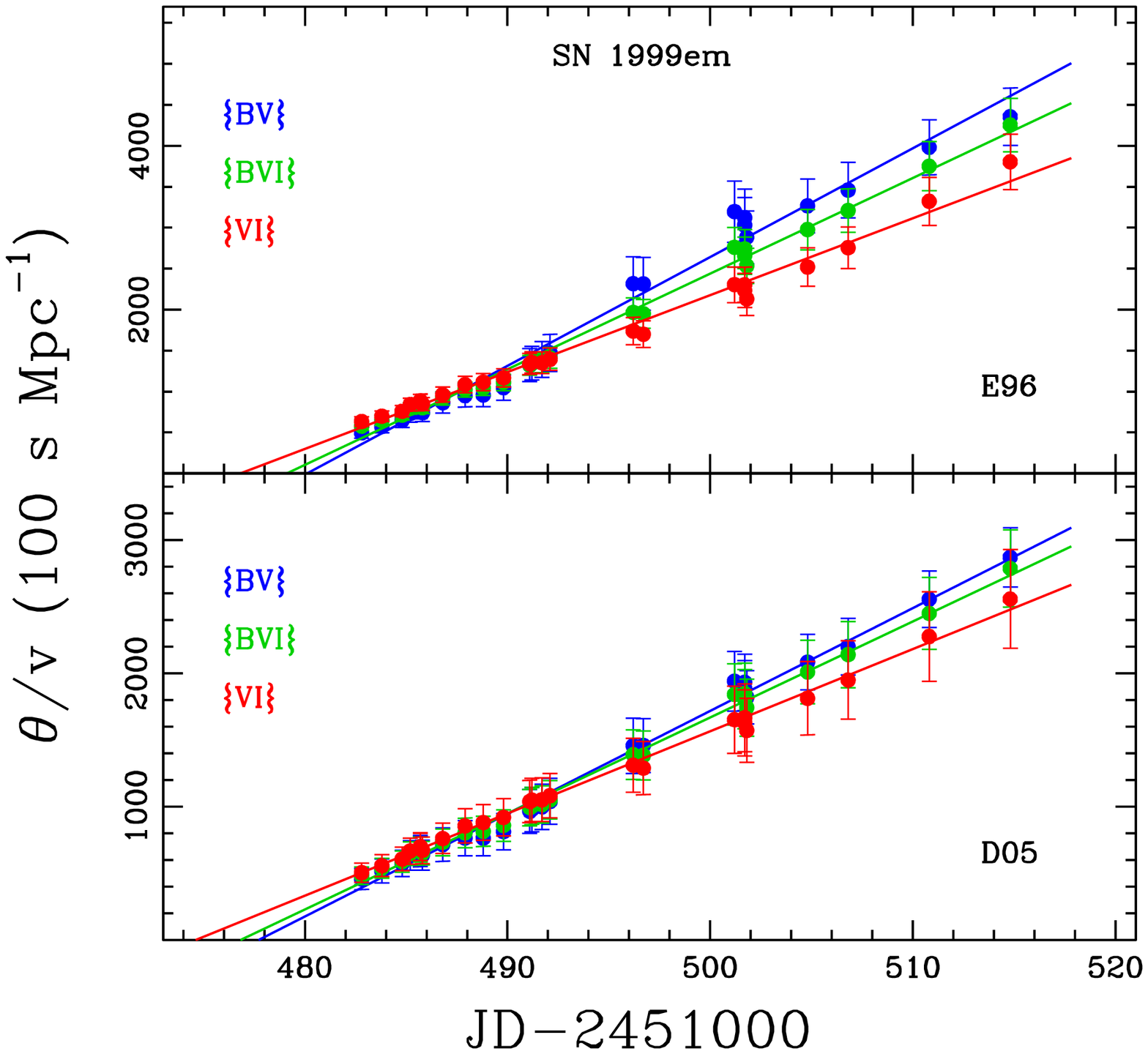}
\caption{The ratio $\theta/v$ as a function of time for SN 1999em
using the $\{BV\}$, $\{BVI\}$, and $\{VI\}$ filter subsets. The ridge
lines correspond to unweighted least-squares fits to the derived EPM
quantities. The upper and lower panels show the results using {\rm
E96} and {\rm D05} dilution factors, respectively. In all cases we
employ the {\rm DES} reddening.
~\label{fig_SN99em_EPM}}
\end{figure}

\begin{figure}
\plotone{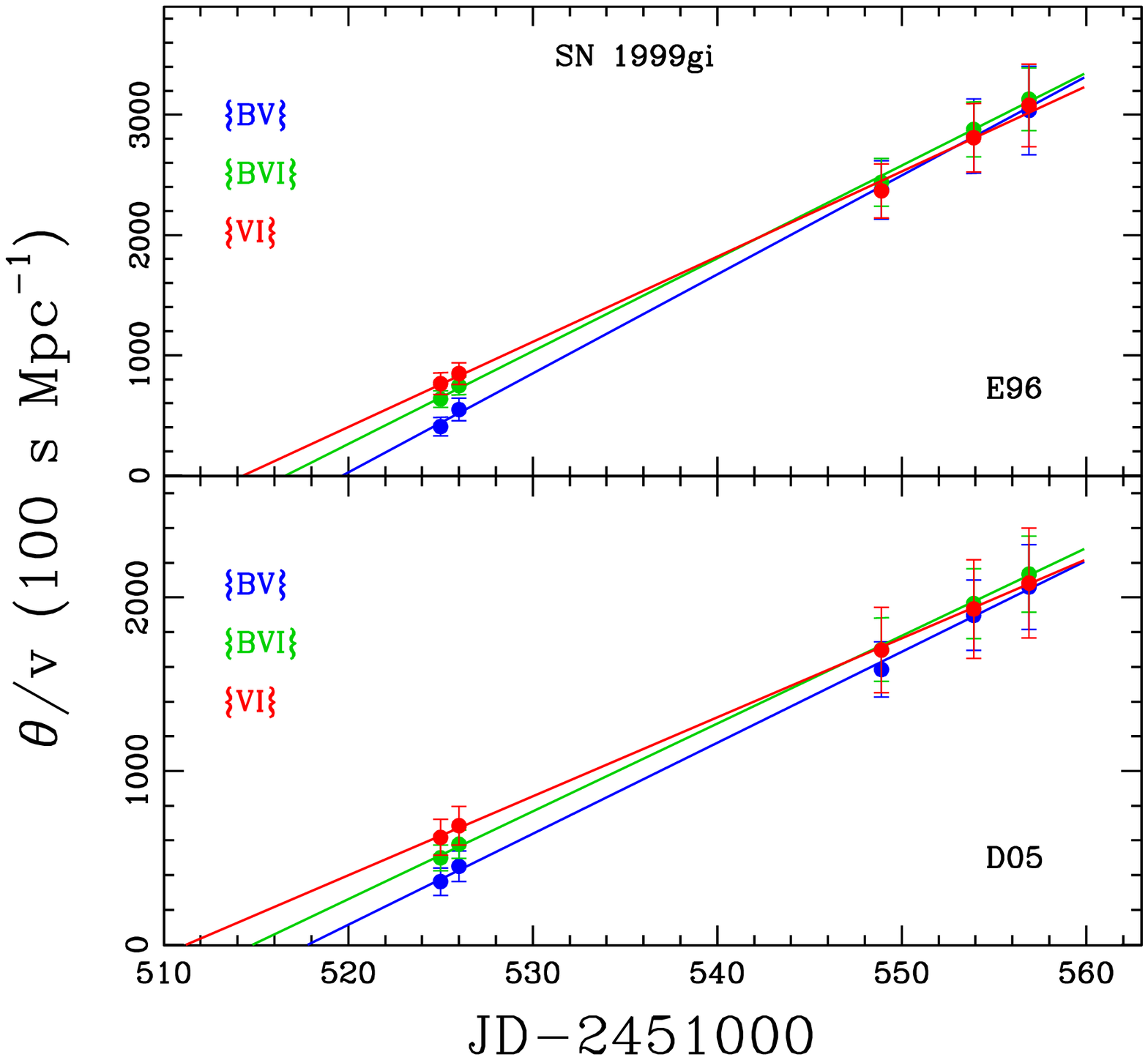}
\caption{The ratio $\theta/v$ as a function of time for SN 1999gi
using the $\{BV\}$, $\{BVI\}$, and $\{VI\}$ filter subsets. The ridge
lines correspond to unweighted least-squares fits to the derived EPM
quantities. The upper and lower panels show the results using {\rm
E96} and {\rm D05} dilution factors, respectively. In all cases we
employ the {\rm DES} reddening.
~\label{fig_SN99gi_EPM}}
\end{figure}

\begin{figure}
\plotone{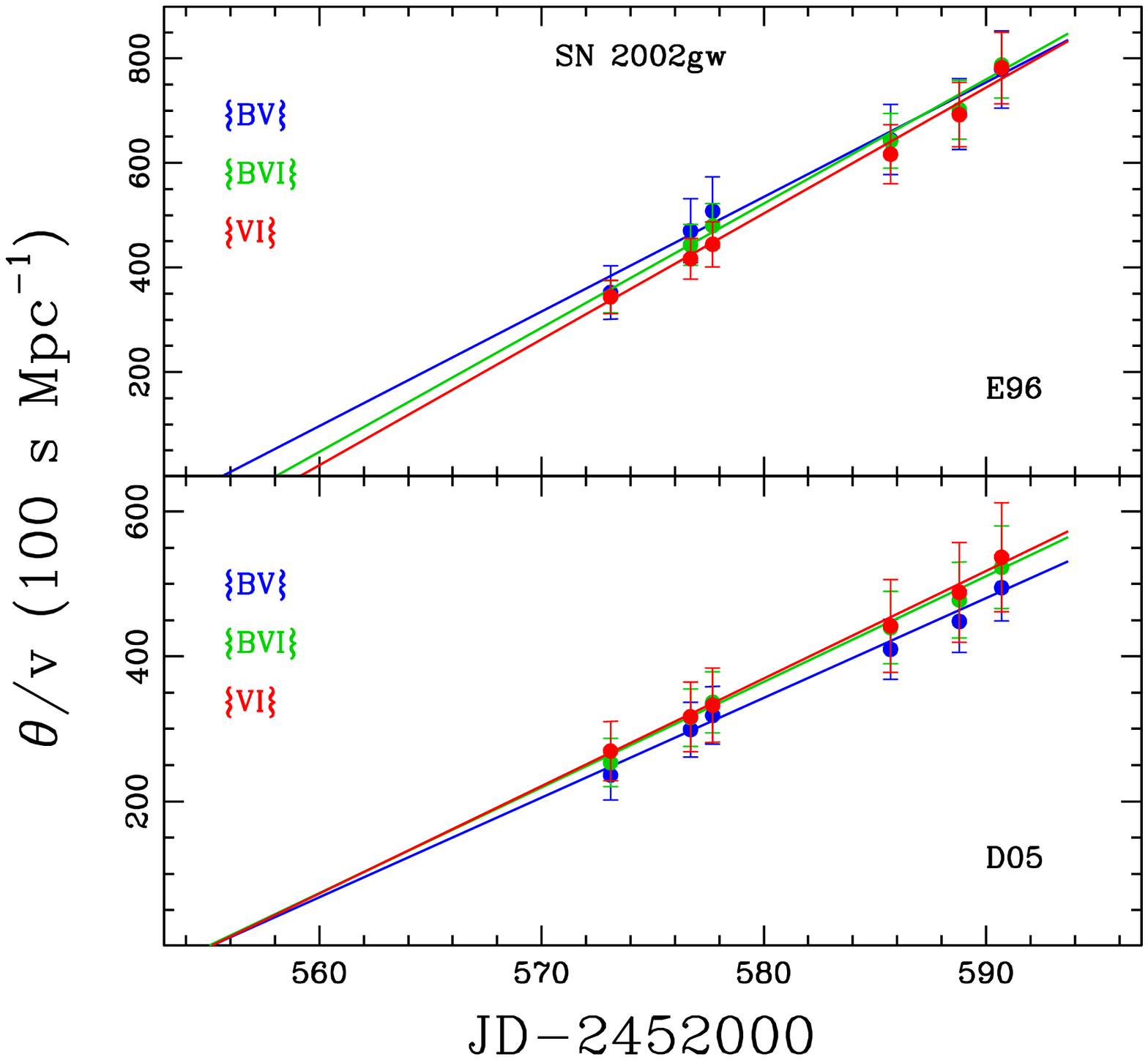}
\caption{The ratio $\theta/v$ as a function of time for SN 2002gw
using the $\{BV\}$, $\{BVI\}$, and $\{VI\}$ filter subsets. The ridge
lines correspond to unweighted least-squares fits to the derived EPM
quantities. The upper and lower panels show the results using {\rm
E96} and {\rm D05} dilution factors, respectively. In all cases we
employ the {\rm DES} reddening.
~\label{fig_SN02gw_EPM}}
\end{figure}

\begin{figure}
\plotone{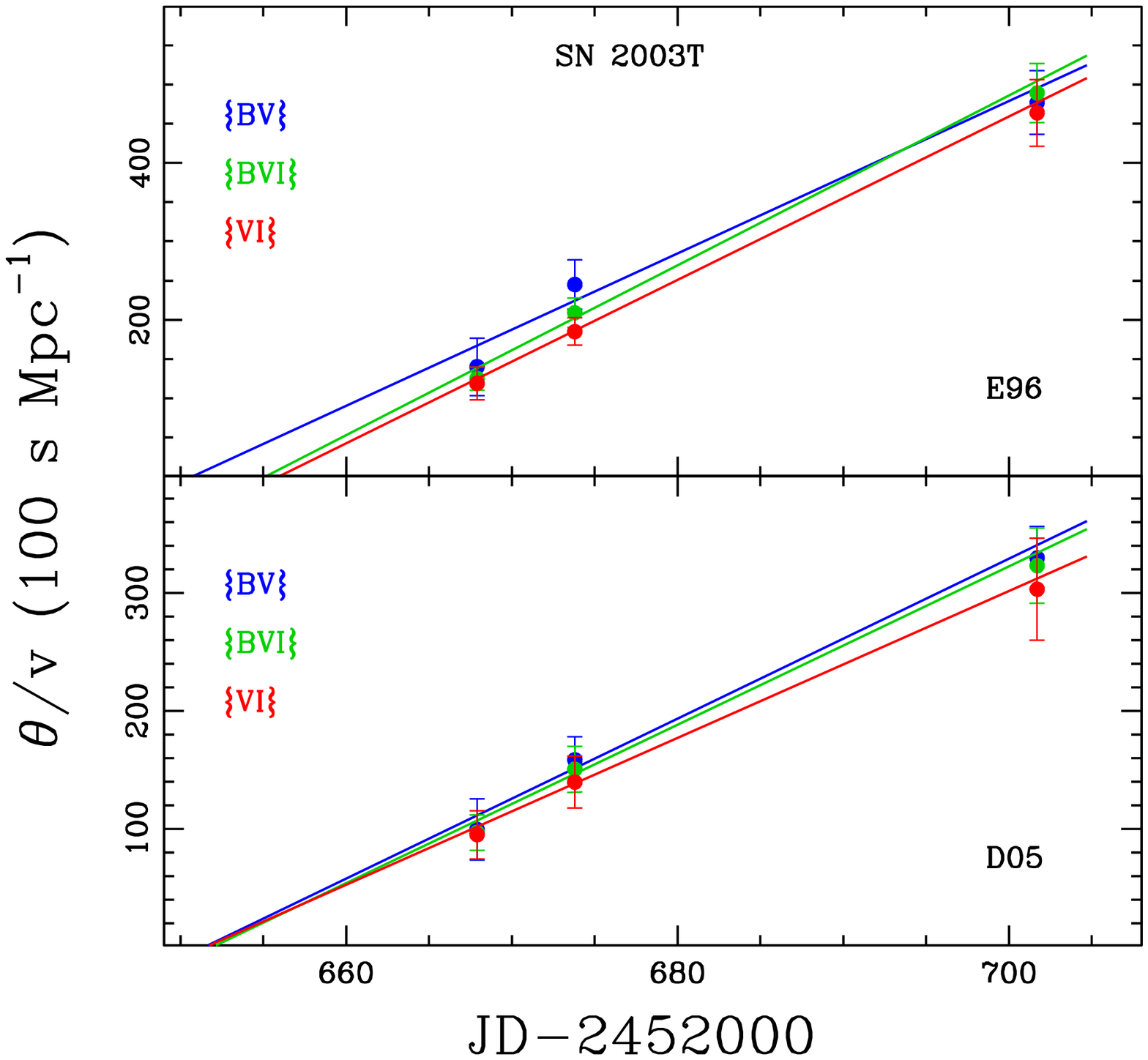}
\caption{The ratio $\theta/v$ as a function of time for SN 2003T using
the $\{BV\}$, $\{BVI\}$, and $\{VI\}$ filter subsets. The ridge lines
correspond to unweighted least-squares fits to the derived EPM
quantities. The upper and lower panels show the results using {\rm
E96} and {\rm D05} dilution factors, respectively. In all cases we
employ the {\rm DES} reddening.
~\label{fig_SN03T_EPM}}
\end{figure}

\begin{figure}
\plotone{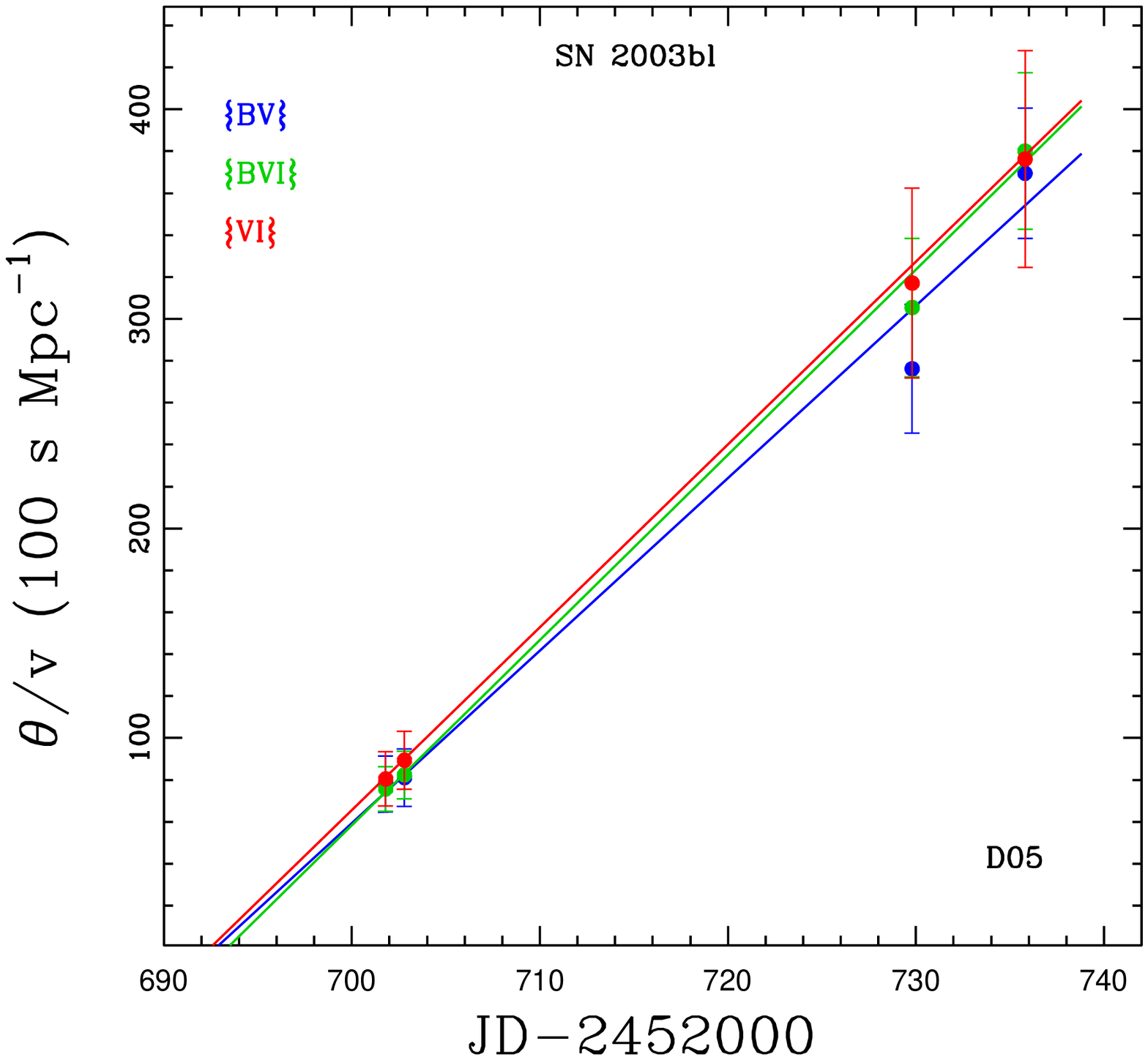}
\caption{The ratio $\theta/v$ as a function of time for SN 2003bl
using the $\{BV\}$, $\{BVI\}$, and $\{VI\}$ filter subsets and the {\rm
D05} models. The ridge lines correspond to unweighted least-squares
fits to the derived EPM quantities. In all cases we employ the {\rm
DES} reddening.
~\label{fig_SN03bl_EPM}}
\end{figure}

\begin{figure}
\plotone{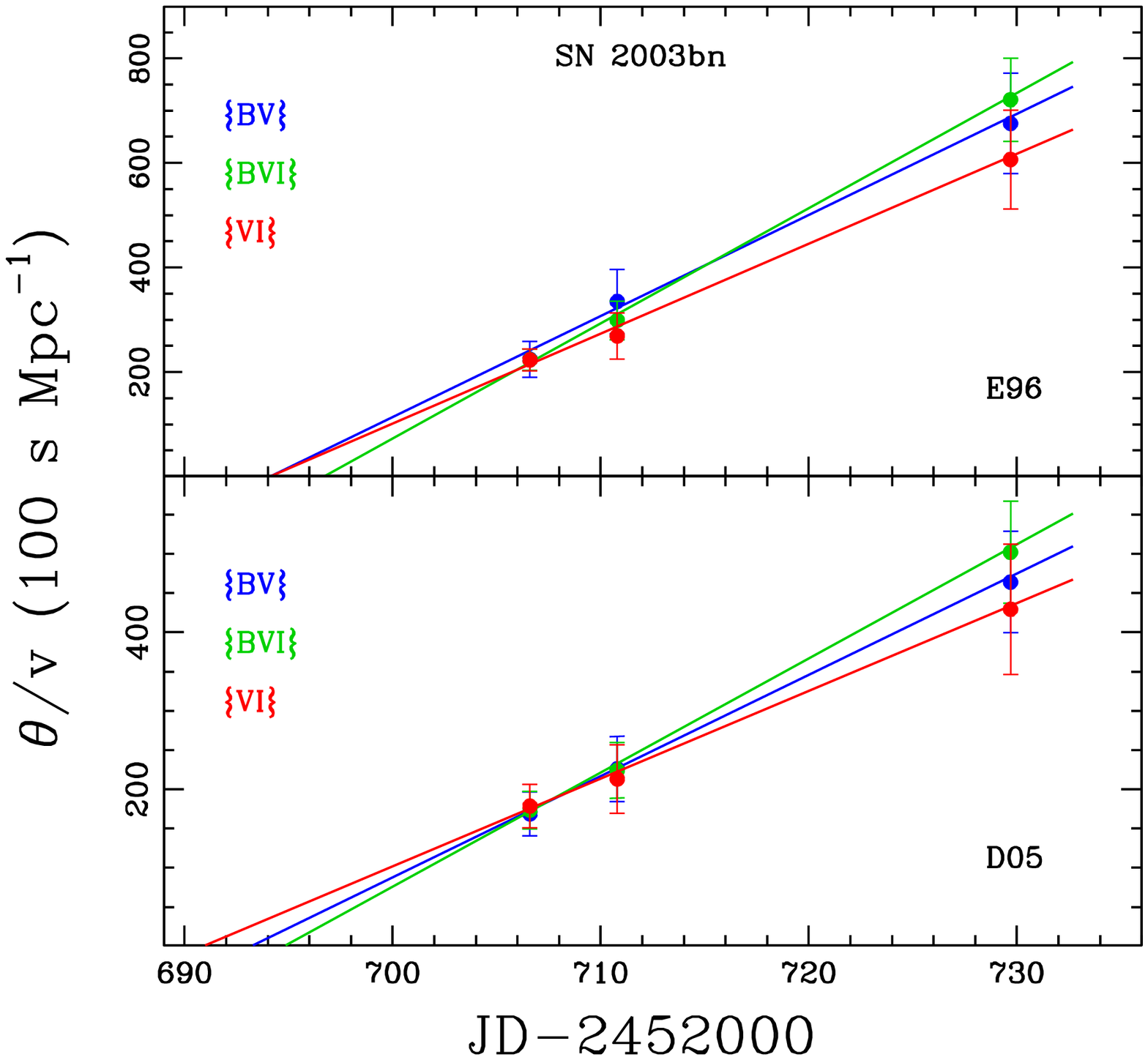}
\caption{The ratio $\theta/v$ as a function of time for SN 2003bn
using the $\{BV\}$, $\{BVI\}$, and $\{VI\}$ filter subsets. The ridge
lines correspond to unweighted least-squares fits to the derived EPM
quantities. The upper and lower panels show the results using {\rm
E96} and {\rm D05} dilution factors, respectively. In all cases we
employ the {\rm DES} reddening.
~\label{fig_SN03bn_EPM}}
\end{figure}

\begin{figure}
\plotone{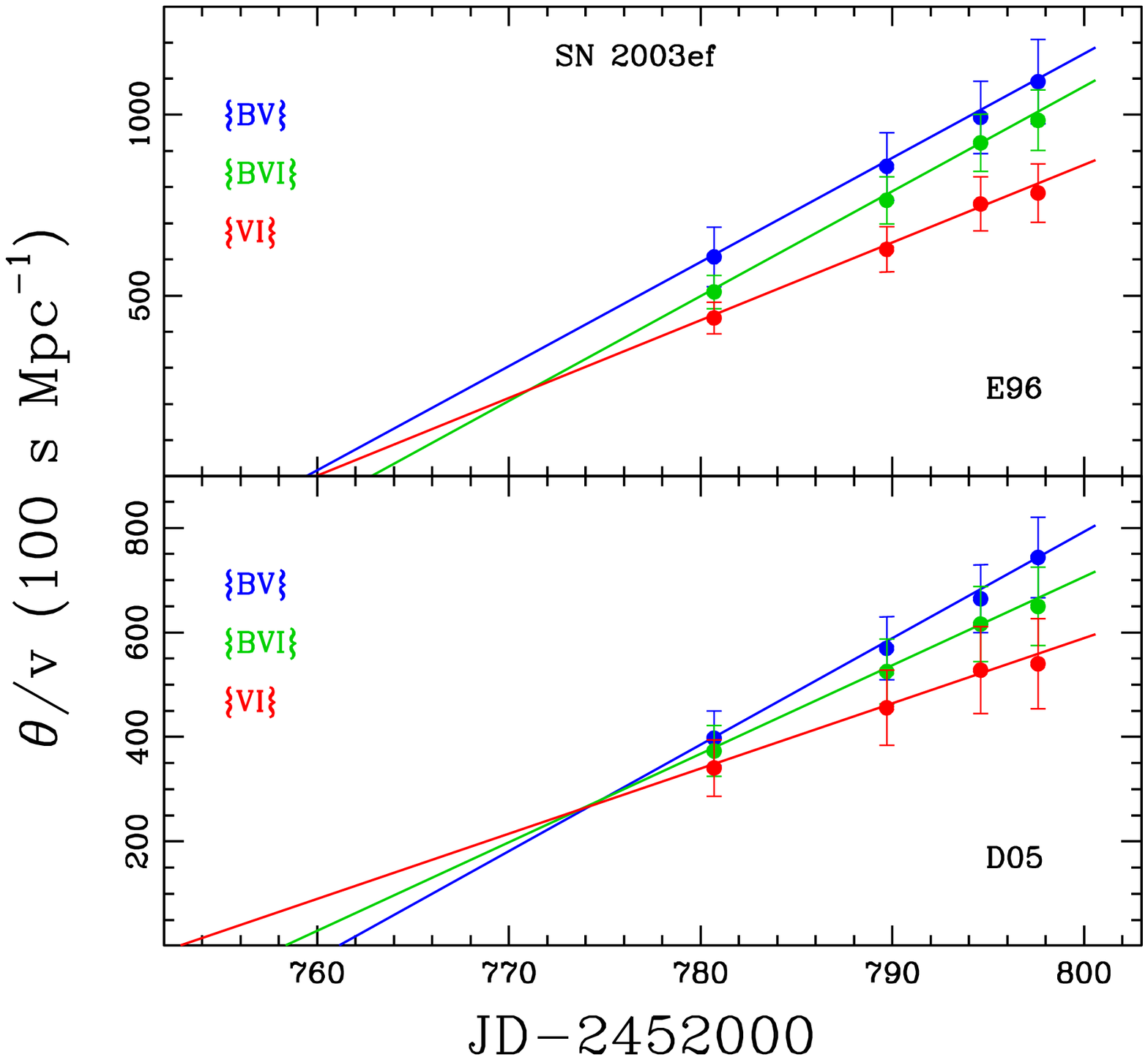}
\caption{The ratio $\theta/v$ as a function of time for SN 2003ef
using the $\{BV\}$, $\{BVI\}$, and $\{VI\}$ filter subsets. The ridge
lines correspond to unweighted least-squares fits to the derived EPM
quantities. The upper and lower panels show the results using {\rm
E96} and {\rm D05} dilution factors, respectively. In all cases we
employ the {\rm DES} reddening.
~\label{fig_SN03ef_EPM}}
\end{figure}

\begin{figure}
\plotone{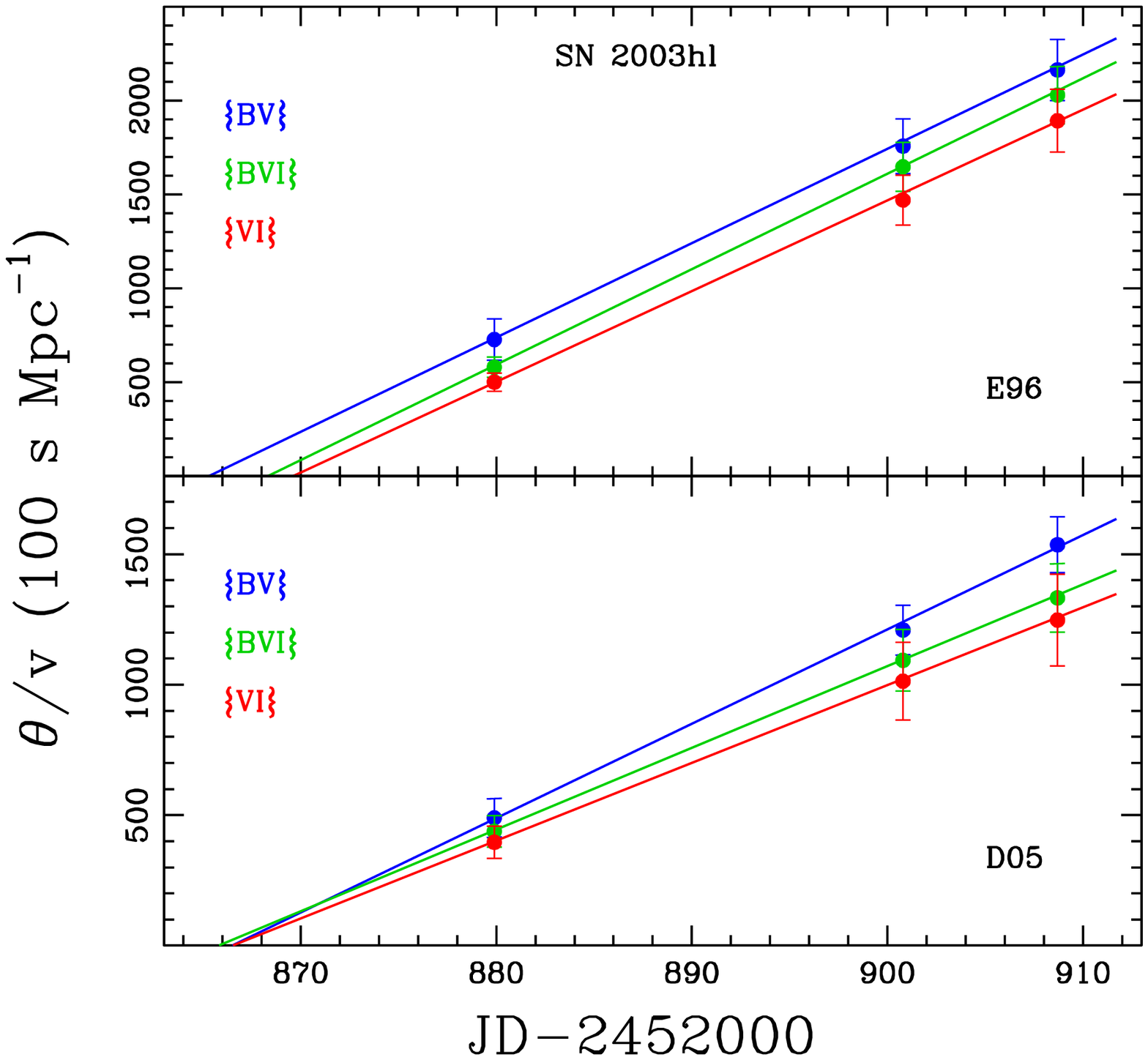}
\caption{The ratio $\theta/v$ as a function of time for SN 2003hl
using the $\{BV\}$, $\{BVI\}$, and $\{VI\}$ filter subsets. The ridge
lines correspond to unweighted least-squares fits to the derived EPM
quantities. The upper and lower panels show the results using {\rm
E96} and {\rm D05} dilution factors, respectively. In all cases we
employ the {\rm DES} reddening.
~\label{fig_SN03hl_EPM}}
\end{figure}

\begin{figure}
\plotone{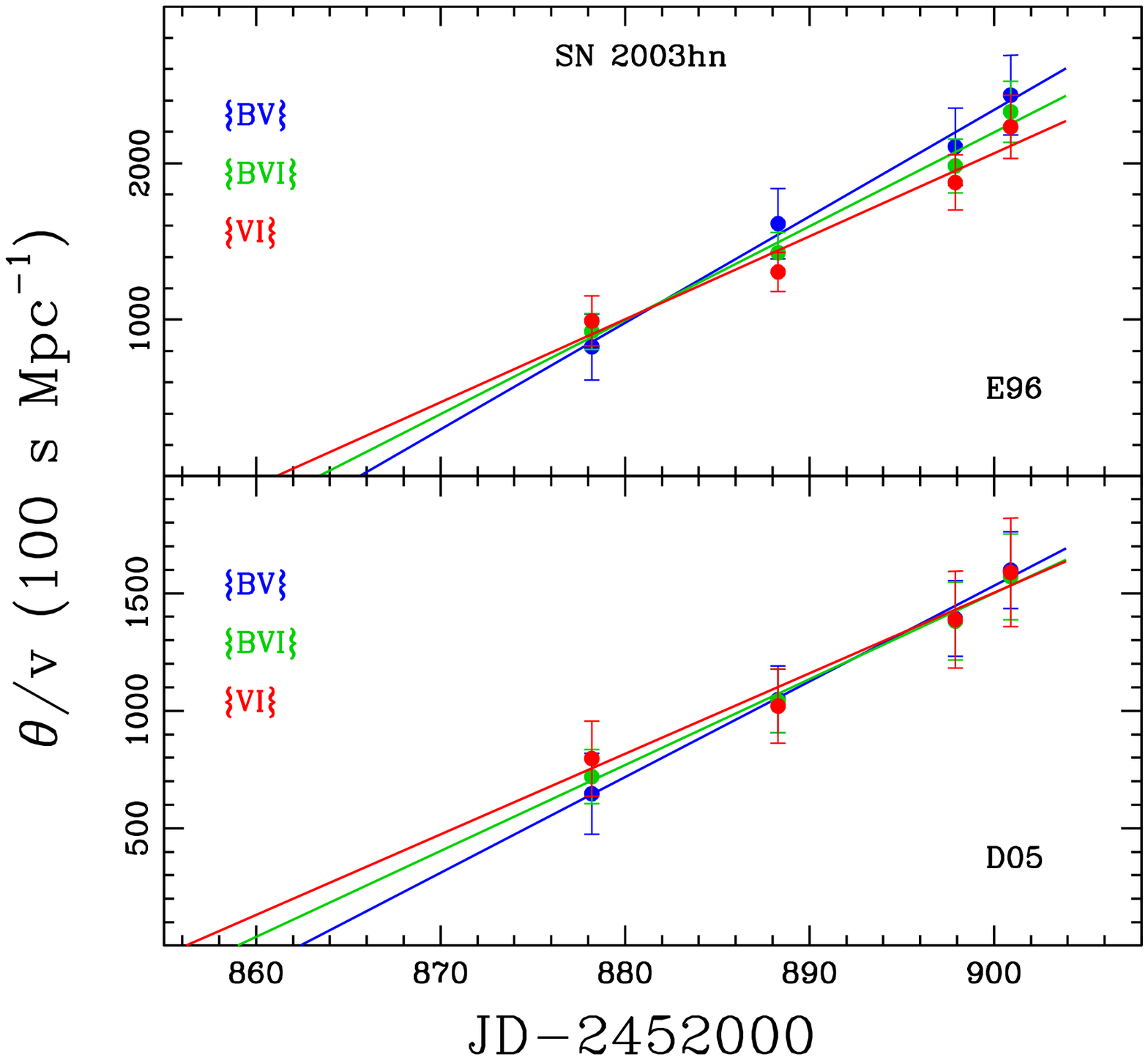}
\caption{The ratio $\theta/v$ as a function of time for SN 2003hn
using the $\{BV\}$, $\{BVI\}$, and $\{VI\}$ filter subsets. The ridge
lines correspond to unweighted least-squares fits to the derived EPM
quantities. The upper and lower panels show the results using {\rm
E96} and {\rm D05} dilution factors, respectively. In all cases we
employ the {\rm DES} reddening.
~\label{fig_SN03hn_EPM}}
\end{figure}

\begin{figure}
\plotone{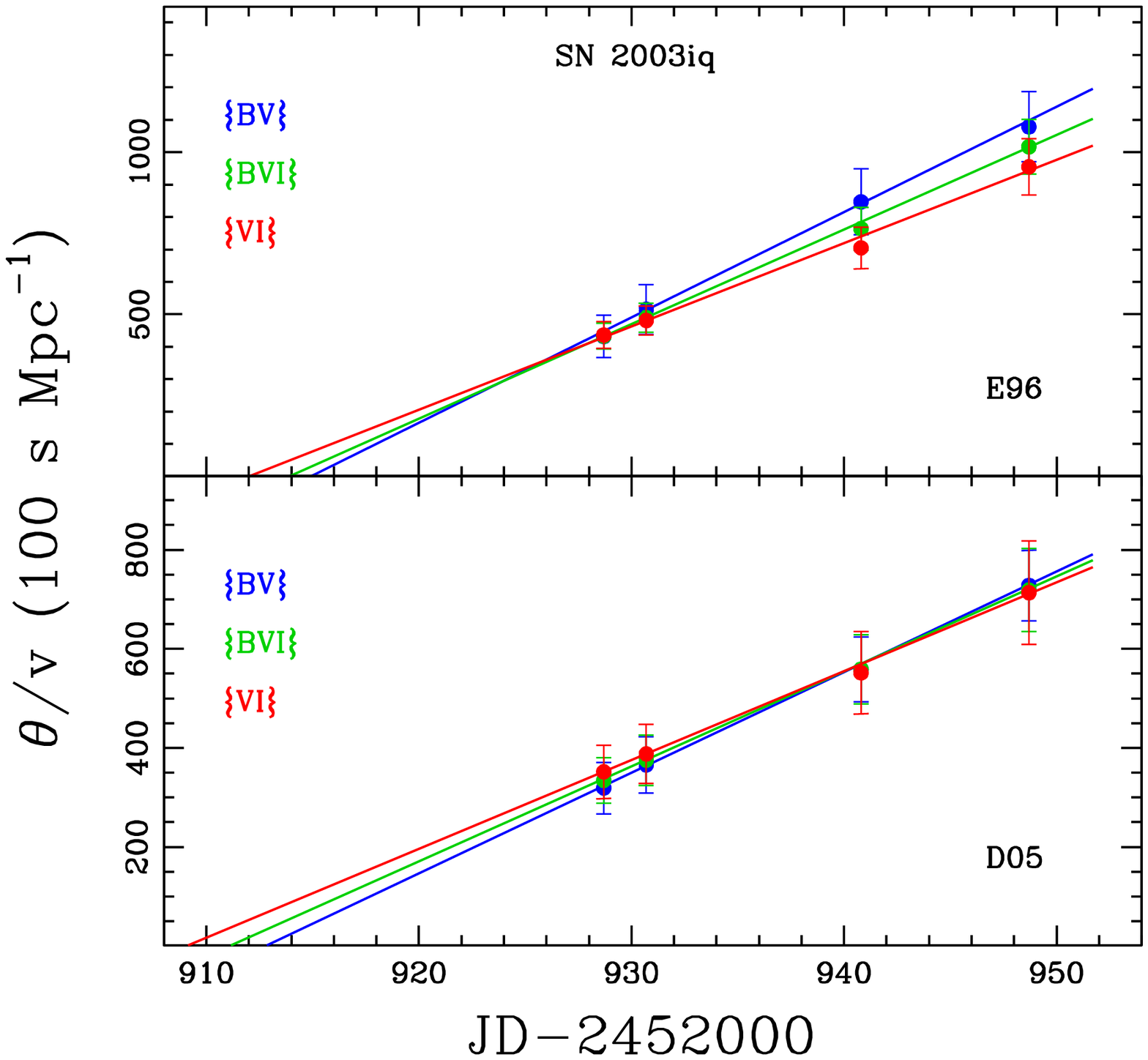}
\caption{The ratio $\theta/v$ as a function of time for SN 2003iq
using the $\{BV\}$, $\{BVI\}$, and $\{VI\}$ filter subsets. The ridge
lines correspond to unweighted least-squares fits to the derived EPM
quantities. The upper and lower panels show the results using {\rm
E96} and {\rm D05} dilution factors, respectively. In all cases we
employ the {\rm DES} reddening.
~\label{fig_SN03iq_EPM}}
\end{figure}

We present here the EPM analysis of 12 SNe~IIP (11 from our database
and one from the literature) with early spectroscopic and photometric
observations. We carried out the analysis using three different
filter subsets ($\{BV\},\, \{BVI\},\, \{VI\}$), two sets of host-galaxy
extinction ({\rm OLI}, {\rm DES}), and two atmosphere models ({\rm
E96}, {\rm D05}), yielding a total of 12 solutions for each SN.
The tables that summarize the EPM quantities are available in
electronic format for all 144 cases.
In the remainder of this section we restrict the presentation to the 6
solutions that use the {\rm DES} extinction because they give the
lowest dispersion in the Hubble diagrams. Figures
\ref{fig_SN92ba_EPM}--\ref{fig_SN03iq_EPM} show these 6 solutions for
each of the 12 SNe. Below, we provide the EPM distance $D$
and the explosion date $t_0$ and their uncertainties, using {\rm DES}
and the \{VI\} filter subset, and we compare the time of explosion to
the range restricted by pre-SN images of the host galaxies. These
results are also summarized in Table \ref{tab_VI_DES.EPM}. 
\newline \indent
In order to obtain a more realistic estimation of the uncertainty in
the distance and the explosion date, we computed 100 Monte Carlo
simulations for each SN, in which we varied all the parameters
involved in the EPM (see Table \ref{tab_errors}), and we averaged the
100 distances and explosion dates to derive the EPM values of $D$ and
$t_0$. This produces small differences between the results computed
from the initial single EPM solution and that obtained from the 100
Monte Carlo simulations, but the latter provides a much more realistic
estimate of the uncertainties.

\subsubsection*{SN 1992ba}

Figure \ref{fig_SN92ba_EPM} shows $\theta/v$ versus time for SN 1992ba
using the $\{BV\},\, \{BVI\}$, and $\{VI\}$ filter subsets and the {\rm
E96} and {\rm D05}.
We used 3 epochs (JD 2448896.9--2448922.8) to compute the distance to
this SN. In order to use the velocities measured on JD 2448896.9 and
2448900.9, we had to extrapolate the $I$-band photometry until JD
2448896.9.
\newline \indent
SN 1992ba was discovered by \citet{Eva92} on JD
2448896.3. \citet{McN92} reported that the SN was not present on a
plate taken on JD 2448883.2 with limiting magnitude 19. The EPM
solution yields $t_{0}$ = JD 2448883.9 $\pm$ 3.0 using the {\rm E96}
models and $t_{0}$ = JD 2448879.8 $\pm$ 5.6 with {\rm D05}. These
results agree (within 1$\sigma$) with the explosion date constrained
by the pre- and post-explosion observations. The distances derived to
SN 1992ba are $D$ = 16.4 $\pm$ 2.5 Mpc and $D$ = 27.2 $\pm$ 6.5 Mpc
using the {\rm E96} and the {\rm D05} dilution factors, respectively.

\subsubsection*{SN 1999br}

Figure \ref{fig_SN99br_EPM} shows $\theta/v$ versus time for SN 1999br
using the $\{BV\},\, \{BVI\}$, and $\{VI\}$ filter subsets and the {\rm
D05} atmosphere models.
We used 5 epochs (JD 2451291.7--2451309.7) to compute the distance to
this SN. The EPM solution shows some departure from linearity using
the \{BV\} and \{BVI\} filter subsets. SN 1999br exhibits very low
expansion velocities, so we were unable to obtain its distance using
the {\rm E96} models. This is because the photospheric velocity
conversion factor $V_{H\beta}/V_{phot}$ is not defined at low
expansion velocities (see $\S$ 3.4 and Figure
\ref{fig_models_VHbeta_Vphot}). The EPM solution yields $t_{0}$ = JD
2451275.6 $\pm$ 7.7 using the {\rm D05} models. This result compares
very well with the observations, because SN 1999br was discovered by
the Lick Observatory Supernova Search \citep[LOSS;][]{fil01}
with the Katzman Automatic Imaging Telescope (KAIT) on JD 2451280.9
\citep{Kin99}. An image taken on JD 2451264.9 showed nothing at the
SN position at a limiting magnitudes of 18.5 \citep{Li99a}. The EPM
distance to SN 1999br is $D$ = 39.5 $\pm$ 13.5 Mpc using the {\rm D05}
dilution factors.

\subsubsection*{SN 1999em}

SN 1999em is the best-ever observed SN IIP. Many photometric and
spectroscopic observations were made by different observers during the
plateau phase. Figure \ref{fig_SN99em_EPM} shows $\theta/v$ versus
time for the SN 1999em using the $\{BV\},\, \{BVI\}$, and $\{VI\}$
filter subsets and the {\rm E96} and {\rm D05} models. Table
\ref{tab_SN99em_EPM} summarizes the EPM quantities derived from the
$\{VI\}$ filter subset. We used 25 epochs (JD 2451482.8--2451514.8) to
derive the distance to SN 1999em. Four spectra were taken from
\citet{Ham01} and the other 21 from \citet{Leo02b}. In some cases
there were two spectra taken at the same epoch from both sources; we
used them individualy in the EPM solution instead of averaging the
measured velocities from each spectrum. We removed the first spectrum
(JD 2451481.8) from the EPM solution because it shows a clear
departure from the linear $\theta/v$ versus $t$ relation. The EPM
solutions using the {\rm E96} and {\rm D05} models are quite linear
and show great detail in the evolution of $\theta/v$ due to the
high-quality spectroscopic and photometric coverage. However, the
{\rm E96} solution shows a small departure from linearity in the last
two spectroscopic epochs. This effect is probably due to the high rise
in the $V_{H\beta} /V_{phot}$ ratio at low velocities in the {\rm E96}
models.
\newline \indent
 SN 1999em was discovered on JD 2451480.9 by the LOSS
\citep{Li99b}. An image taken at the position of the SN on JD
2451472.0 showed nothing at a limiting magnitude of 19.0. The EPM
yields $t_{0}$ = JD 2451476.3 $\pm$ 1.1 and $t_{0}$ = JD 2451474.0
$\pm$ 2.0 using the {\rm E96} and {\rm D05} models, respectively.
These explosions dates are between the pre-discovery and discovery
dates. The distances derived to SN 1999em are $D$ = 9.3 $\pm$ 0.5 Mpc
from {\rm E96} and $D$ = 13.9 $\pm$ 1.4 Mpc from {\rm D05}.

\subsubsection*{SN 1999gi}

Figure \ref{fig_SN99gi_EPM} shows $\theta/v$ versus time for SN 1999gi
using the $\{BV\},\, \{BVI\}$, and $\{VI\}$ filter subsets and the {\rm
E96} and {\rm D05} models.
We used 5 epochs (JD 2451525.0--2451556.9) to apply the EPM
method. All of the spectra and the photometry were taken from
\citet{Leo02a}. The first spcetrum (JD 2451522.9) was removed from the
EPM solutions because it yields an H$\beta$ velocity of ${\rm \sim\,
26000~km~s^{-1}}$, well above the range of the photospheric velocity
conversion (see $\S$ 3.4 and Figure \ref{fig_models_VHbeta_Vphot}).
The explosion dates of SN 1999gi obtained using the EPM are $t_{0}$ =
JD 2451517.0 $\pm$ 1.2 using the {\rm E96} models and $t_{0}$ = JD
2451515.6 $\pm$ 2.4 with {\rm D05}. These results agree with the
observations because a pre-discovery image taken on JD 2451515.7
\citep{Tr99} showed nothing at the SN position (limiting unfiltered
magnitude of 18.5). SN 1999gi was discovered on JD 2451522.3
\citep{Na99} on unfiltered CCD frames, so the explosion date can be
constrained in a range of only 6.6 days. We derive a distance of $D$
= 11.7 $\pm$ 0.8 Mpc and $D$ = 17.4 $\pm$ 2.3 Mpc using the {\rm E96}
and {\rm D05} models, respectively.

\subsubsection*{SN 2002gw}

Figure \ref{fig_SN02gw_EPM} shows $\theta/v$ versus time for SN 2002gw
using the $\{BV\},\, \{BVI\}$, and $\{VI\}$ filter subsets and the {\rm
E96} and {\rm D05} models.
The EPM solutions were obtained using 6 epochs (JD
2452573.1--2452590.7). The EPM yields explosion times of $t_{0}$ = JD
2452557.9 $\pm$ 2.7 and $t_{0}$ = JD 2452551.7 $\pm$ 7.6 (using {\rm
  E96} and {\rm D05} dilution factors, respectively). SN 2002gw was
discovered on JD 2452560.8 \citep{Mon02}. An image taken on JD
2452529.6 shows nothing at the SN position at a limiting magnitude of
18.5. Also, an unfiltered CCD image taken on JD 2452559.1 shows the SN
at magnitude 18.3 \citep{Ita02}. The EPM explosion dates are in
agreement with the SN explosion date constrained by the
observations. The EPM distances are $D$ = 37.4 $\pm$ 4.9 Mpc and $D$ =
63.9 $\pm$ 17.0 Mpc using {\rm E96} and {\rm D05}, respectively.

\subsubsection*{SN 2003T}

Figure \ref{fig_SN03T_EPM} shows $\theta/v$ versus time for SN 2003T
using the $\{BV\},\, \{BVI\}$ and $\{VI\}$ filter subsets and the {\rm
E96} and {\rm D05} models.
The EPM solutions were obtained using 3 epochs (JD 2452667.9-2452701.7).
The EPM explosion dates are $t_{0}$ = JD 2452654.2 $\pm$ using {\rm
E96} models and $t_{0}$ = JD 2452648.9 $\pm$ 3.4 with {\rm D05}. In
both cases the third epoch used to derive the distance is beyond ${\rm
\sim\,45}$ days after the EPM $t_{0}$, but it proves neccesary to
include it to compute the EPM analysis. This SN was discovered by the
Lick and Tenagra Observatories Supernova Search (LOTOSS) on JD
2452664.9 \citep{Sch03}. An image taken on JD 2452644.9 shows nothing
at a limiting magnitude of 19.0, in good agreement with the EPM
analysis. The EPM distances are $D$ = 87.8 $\pm$ 13.5 Mpc using {\rm
E96} and $D$ = 147.3 $\pm$ 35.7 Mpc with {\rm D05}.

\subsubsection*{SN 2003bl}

Figure \ref{fig_SN03bl_EPM} shows $\theta/v$ versus time for SN 2003bl
using the $\{BV\},\, \{BVI\}$, and $\{VI\}$ filter subsets and {\rm D05}
models.
The EPM solutions were obtained using 4 epochs (JD
2452701.8--2452735.8). As with the SN 1999br, we were unable to apply
the EPM using {\rm E96} because we only had two spectra with
velocities higher than ${\rm 4500~km~s^{-1}}$, and so the photospheric
velocity correction could not be applied (see $\S$ 3.4 and Figure
\ref{fig_models_VHbeta_Vphot}). SN 2003bl was discovered by LOTOSS on
JD 2452701.0 \citep{Swi03}. A pre-discovery image taken on JD
2452438.8 shows nothing at the SN position at a limiting magnitud of
19.0. The EPM yields $t_{0}$ = JD 2452692.6 $\pm$ 2.8, consistent
with the SN discovery date. The EPM distance is $D$ = 92.4 $\pm$
14.2 Mpc.

\subsubsection*{SN 2003bn}

Figure \ref{fig_SN03bn_EPM} shows $\theta/v$ versus time for SN 2003bn
using the $\{BV\},\, \{BVI\}$, and $\{VI\}$ filter subsets and the {\rm
E96} and {\rm D05} models.
We computed the EPM analysis using 3 epochs (JD 2452706.6--2452729.7).
The EPM yields explosions dates of $t_{0}$ = JD 2452693.4 $\pm$ 2.7
and $t_{0}$ = JD 2452687.0 $\pm$ 9.0 from {\rm E96} and {\rm D05},
respectively. SN 2003bn was discovered on JD 2452698.0
\citep{Wood03}. Two pre-discovery NEAT images shows nothing at the SN
position on JD 2452691.5 (limiting magnitude of 21.0) and the SN at a
magnitude of 20.2 on JD 2452692.8, which restricted the explosion date
in a range of only 1.3 days. This value for $t_0$ is in agreement
within one $\sigma$ with the EPM $t_0$ derived using {\rm E96} and
{\rm D05}. The EPM distances from {\rm E96} and {\rm D05} are $D$ =
50.2 $\pm$ 7.0 Mpc and $D$ = 87.2 $\pm$ 28.0 Mpc, respectively.

\subsubsection*{SN 2003ef}

Figure \ref{fig_SN03ef_EPM} shows $\theta/v$ versus time for SN 2003ef
using the $\{BV\},\, \{BVI\}$, and $\{VI\}$ filter subsets and the {\rm
E96} and {\rm D05} models.
We computed the EPM analysis using 4 epochs (JD 2452780.7--2452797.6).
The explosion date derived are $t_{0}$ = JD 2452759.8 $\pm$ 4.7 and
$t_{0}$ = JD 2452748.4 $\pm$ 15.6 with {\rm E96} and {\rm D05},
respectively. SN 2003ef was discovered by the LOTOSS on JD 2452770.8
(mag. about 16.3) \citep{Wei03}, consistent with the EPM value of
$t_{0}$. A KAIT image taken on JD 2452720.8 showed nothing at the SN
position at a limiting magnitude of 18.5. The EPM distances are $D$ =
38.7 $\pm$ 6.53 Mpc with {\rm E96} and $D$ = 74.4 $\pm$ 30.3 Mpc with
{\rm D05}.
 
\subsubsection*{SN 2003hl}

Figure \ref{fig_SN03hl_EPM} shows $\theta/v$ versus time for SN 2003hl
using the $\{BV\},\, \{BVI\}$, and $\{VI\}$ filter subsets and the {\rm
E96} and {\rm D05} models.
The EPM solutions were obtained using 3 epochs (JD
2452879.9--2452908.7). We estimated the explosion dates on $t_{0}$ = JD
2452872.3 $\pm$ 1.7 and $t_{0}$ = JD 2452865.4 $\pm$ 5.9 using {\rm E96}
and {\rm D05}, respectively. SN 2003hl was discovered on JD 2452872.0
during the LOTOSS program at a magnitude of 16.5 \citep{Moo03}. A
pre-discovery KAIT image taken on JD 2452863.0 shows nothing at the SN
position at a limiting magnitude of 19.0. This image restricts the
explosion date in a range of 9 days. The EPM explosion dates are in
agreement with the observations (within one $\sigma$). We derived EPM
distances of $D$ = 17.7 $\pm$ 2.1 Mpc with {\rm E96} and $D$ = 30.3
$\pm$ 6.3 Mpc with {\rm D05}.

\subsubsection*{SN 2003hn}

Figure \ref{fig_SN03hn_EPM} shows $\theta/v$ versus time for SN 2003hn
using the $\{BV\},\, \{BVI\}$, and $\{VI\}$ filter subsets and the {\rm
E96} and {\rm D05} models.
The EPM solutions were obtained using 4 epochs (JD
2452878.2--2452900.9). The EPM explosion dates derived are $t_{0}$ =
JD 2452859.5 $\pm$ 3.8 and $t_{0}$ = JD 2452853.8 $\pm$ 9.3 using the
{\rm E96} and {\rm D05} dilution factors, respectively. This SN was
discovered on JD 2452877.2 at 14.1 mag by \citet{Eva03}. Evans also
reported that the SN was not visible at 15.5 mag on JD 2452856.5. This
date agrees with the explosion date derived from {\rm E96} and is $<
1\sigma$ lower than that derived from {\rm D05}. The EPM solutions
lead to $D$ = 16.9 $\pm$ 2.2 Mpc and $D$ = 26.3 $\pm$ 7.1 Mpc using
{\rm E96} and {\rm D05} models, respectively.

\subsubsection*{SN 2003iq}

Figure \ref{fig_SN03iq_EPM} shows $\theta/v$ versus time for SN 2003iq
using the $\{BV\},\, \{BVI\}$, and $\{VI\}$ filter subsets and the {\rm
E96} and {\rm D05} models.
The EPM solutions were obtained using 4 epochs (JD
2452928.7--2452948.7). This SN was discovered by \citet{LLa03} on JD
2452921.5, while monitoring SN 2003hl in the same host galaxy. A
pre-discovery image taken on 2452918.5 shows nothing at the SN
position. These reports constrain the explosion date to a range of
only three days. The EPM method yields $t_{0}$ = JD 2452909.6 $\pm$
4.3 using {\rm E96} and $t_{0}$ = JD 2452905.6 $\pm$ 9.5 using {\rm
  D05}. In both cases the explosion date is far earlier than expected
because the SN was not present on JD 2452918.5. This implies that the
EPM solutions to this SN are not satisfactory. We derived EPM
distances of $D$ = 36.0 $\pm$ 5.6 Mpc with $E96$ and D = 53.3 $\pm$
17.1 Mpc with {\rm D05}.

\section{Discussion}

\subsection{External Comparison}

\subsubsection{Previous EPM distances}

The EPM method has already been applied to SN 1999em by other authors,
as follows. (1) \citet{Ham01} employed the {\rm E96} dilution factors
and eight different filter subsets to perform the EPM analysis to this
SN, using a cross-correlation technique to estimate the photospheric
velocity and adopting a host-galaxy extinction of ${\rm A_V}$ = 0.18 mag.
They derived a distance of 6.9 $\pm$ 0.1, 7.4 $\pm$ 0.1, and 7.3 $\pm$
0.1 Mpc from the $\{BV\}$, $\{BVI\}$, and $\{VI\}$ filter subsets,
respectively. These values are in agreement with our estimates of 6.9
$\pm$ 0.6, 7.5 $\pm$ 0.6, and 9.3 $\pm$ 0.5 Mpc (from the $\{BV\}$,
$\{BVI\}$ and $\{VI\}$ filter subsets, respectively), except in the
$\{VI\}$ case. (2) \citet{Leo02b} employed the {\rm E96} models to
derive the distance to SN 1999em. They used four weak unblended
spectral features (Fe {\sc ii} $\lambda\lambda$\,4629, 5276, 5318, and
Sc {\sc ii} $\lambda$\,4670) as the photospheric velocity indicators.
They adopted a host galaxy reddening of $A_V = 0.31$ mag, the same
value predicted by {\rm DES}. They derived a distance of 7.7 $\pm$
0.2, 8.3 $\pm$ 0.2 and 8.8 $\pm$ 0.3 Mpc from the $\{BV\}$, $\{BVI\}$
and $\{VI\}$ filter subsets, respectively. These results are in
agreement with our {\rm E96} distances. (3) \citet{Elm03} determined a color
temperature by fitting blackbody functions to their observed spectra of
SN 1999em.
They used the {\rm E96} models and adopted a reddening of ${\rm A_V}$ = 0.31 to derive a 
distance of 7.8 $\pm$ 0.3 Mpc to this SN. 
This result is in agreement with our EPM distance using {\rm E96}, except 
in the $\{VI\}$ case.
(4) \citet{D06} applied the EPM method to SN 1999em using {\rm E96} 
and {\rm D05}. They adopted the SN 1999em {\rm DES} reddening value of 
$A_V = 0.31$ mag. Using the {\rm E96} models, they derived a distance of 
8.6 $\pm$ 0.8, 9.7 $\pm$ 1.0, and 11.7 $\pm$ 1.5 Mpc from the $\{BV\}$, $\{BVI\}$, 
and $\{VI\}$ filter subsets, respectively, which are somewhat greater than
our distances. They also used specific {\rm D05} models for this SN
to apply the EPM. They found a distance of 12.4 $\pm$ 1.2, 12.4 $\pm$
1.3, and 12.4 $\pm$ 1.3 Mpc from the $\{BV\}$, $\{BVI\}$, and $\{VI\}$
filter subsets, respectively, using 7 epochs and 11.7 $\pm$ 1.0, 11.6
$\pm$ 1.0 and 11.5 $\pm$ 0.9 Mpc from the $\{BV\}$, $\{BVI\}$, and
$\{VI\}$ filter subsets, respectively, using 8 epochs. These results
compare very well with our distances using {\rm D05} of 11.6 $\pm$
1.2, 12.1 $\pm$ 0.9, and 13.9 $\pm$ 1.4 Mpc from the $\{BV\}$, $\{BVI\}$,
and $\{VI\}$ filter subsets, respectively.

\subsubsection{SEAM Distance}

SEAM is a technique similar to the EPM, but it avoids the use of
dilution factors by doing synthetic spectral fitting to the observed
spectra of the SN. \citet{Bar04} applied this method to SN
1999em. They derived a distance of $D$ = 12.5 $\pm$ 2.3 Mpc, in good
agreement with our distances derived using the {\rm D05} models (11.6
$\pm$ 1.2, 12.1 $\pm$ 0.9, and 13.9 $\pm$ 1.4 Mpc from the $\{BV\}$,
$\{BVI\}$, and $\{VI\}$ filter subsets, respectively), but
significantly greater than the EPM distances derived using {\rm E96}
(6.9 $\pm$ 0.6, 7.5 $\pm$ 0.6 and 9.3 $\pm$ 0.5 Mpc from the $\{BV\}$,
$\{BVI\}$ and $\{VI\}$ filter subsets, respectively). Also \citet{D06}
applied this technique to SN 1999em, deriving a SEAM distance of ${\rm
\sim\,12.2}$ Mpc (11.5 Mpc) using seven (eight) observations, in
agreement with the \citet{Bar04} results.

\subsubsection{Cepheid Distance}

\citet{Leo03} identified 41 Cepheid variable stars in NGC 1637, the
host galaxy of SN 1999em. They derived a Cepheid distance to NGC 1637
of $D$ = 11.7 $\pm$ 1.0 Mpc. As with the SEAM results, the Cepheid
distance is consistent with our EPM distances from the {\rm D05}
models (11.2 $\pm$ 0.2, 12.0 $\pm$ 0.2, and 14.0 $\pm$ 0.2 Mpc from the
$\{BV\}$, $\{BVI\}$, and $\{VI\}$ filter subsets, respectively). In all
cases, the {\rm E96} models lead to significantly lower distances (6.9
$\pm$ 0.6, 7.5 $\pm$ 0.6, and 9.3 $\pm$ 0.5 Mpc from the $\{BV\}$,
$\{BVI\}$, and $\{VI\}$ filter subsets, respectively).

\subsection{Error Analysis}

\subsubsection{Effects of Reddening}

\begin{figure}
\plotone{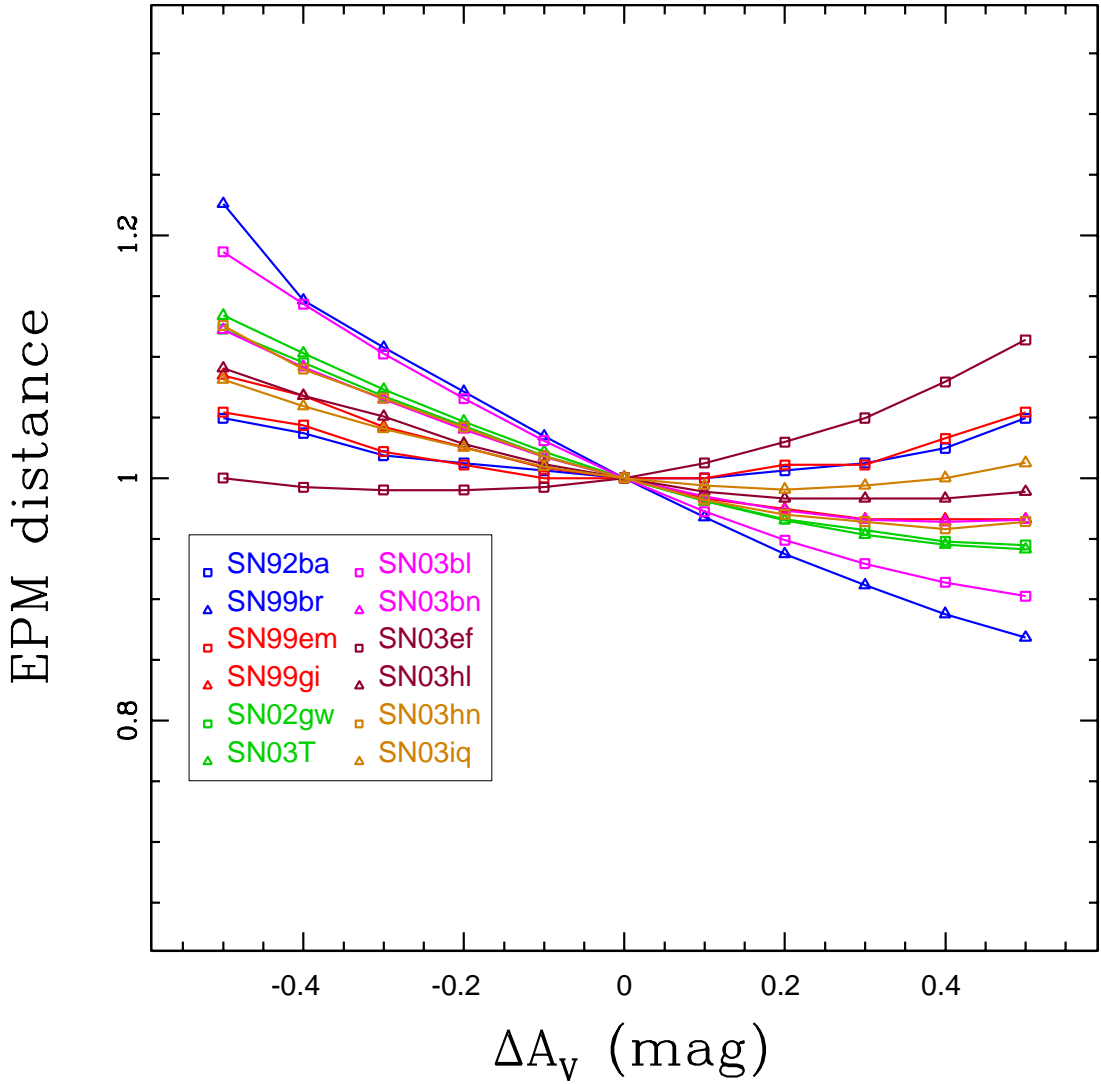}
\caption{Normalized EPM distances as a function of the host-galaxy
visual extinction relative to the {\rm DES} value ($\Delta A_V = 0$).
 ~\label{fig_delta_reddening}}
\end{figure}

While the \citet{Sch98} IR maps provide a reasonable estimate of the
amount of Galactic foreground extinction, the determination of
host-galaxy extinction is a more challenging task. This is a potential
problem because the distances derived using EPM depend on the adopted
host-galaxy extinction. In order to investigate the sensitivity of the
distances to dust extinction, we performed the EPM analysis of all the
SNe in our sample using the $\{VI\}$ filter subset by varying the
amount of host-galaxy visual extinction $A_V$ in steps of $\Delta A_V
= 0.1$ mag. Figure \ref{fig_delta_reddening} shows the normalized EPM
distances as a function of host-galaxy visual extinction $A_V$
relative to the {\rm DES} value (${\rm \Delta A_V\,=\,0}$). As can be
seen, the EPM is quite insensitive to the amount of host galaxy
extinction adopted. On average, the distances change by less than
$\sim\,10$\,\% from $\Delta A_V = 0.0$ to $\Delta A_V = 0.5$ mag and by
less than $\sim\,20$\% from $\Delta A_V = 0.0$ to $\Delta A_V = -0.5$
mag. Therefore, even a large systematic error of 0.5 mag in ${\rm A_V}$
will not translate into a large error in the EPM distance. 
This effect is discussed in \citet{E89}.

\subsubsection{Other Sources of Error}

Table \ref{tab_errors} lists all the error sources in EPM and their
typical values. In order to investigate which source contributes the
most to the uncertainty in the EPM distance, we performed the EPM
analysis of SN 1999gi (whose photometry and spectroscopy coverage is
representative of our sample), and we changed the error of a single
source (listed in Table \ref{tab_errors}) leaving all others
unchanged. We found two main sources of error. In the {\rm E96}
case, the errors in the photospheric velocity conversion and the
dilution factors have the largest effect in the distance uncertainty,
each one contributing $\sim\,30$\,\% of the total error, while
in the {\rm D05} case the error in the dilution factors produces 
$\sim\,70$\,\% of the uncertainty in the distance, far greater than
that due to the error in the photospheric velocity conversion ($\sim\,10$\,\% 
of the total error). All of the other errors have a
secondary effect in the total error.

\subsection{Hubble Diagrams}

\begin{figure}
\plotone{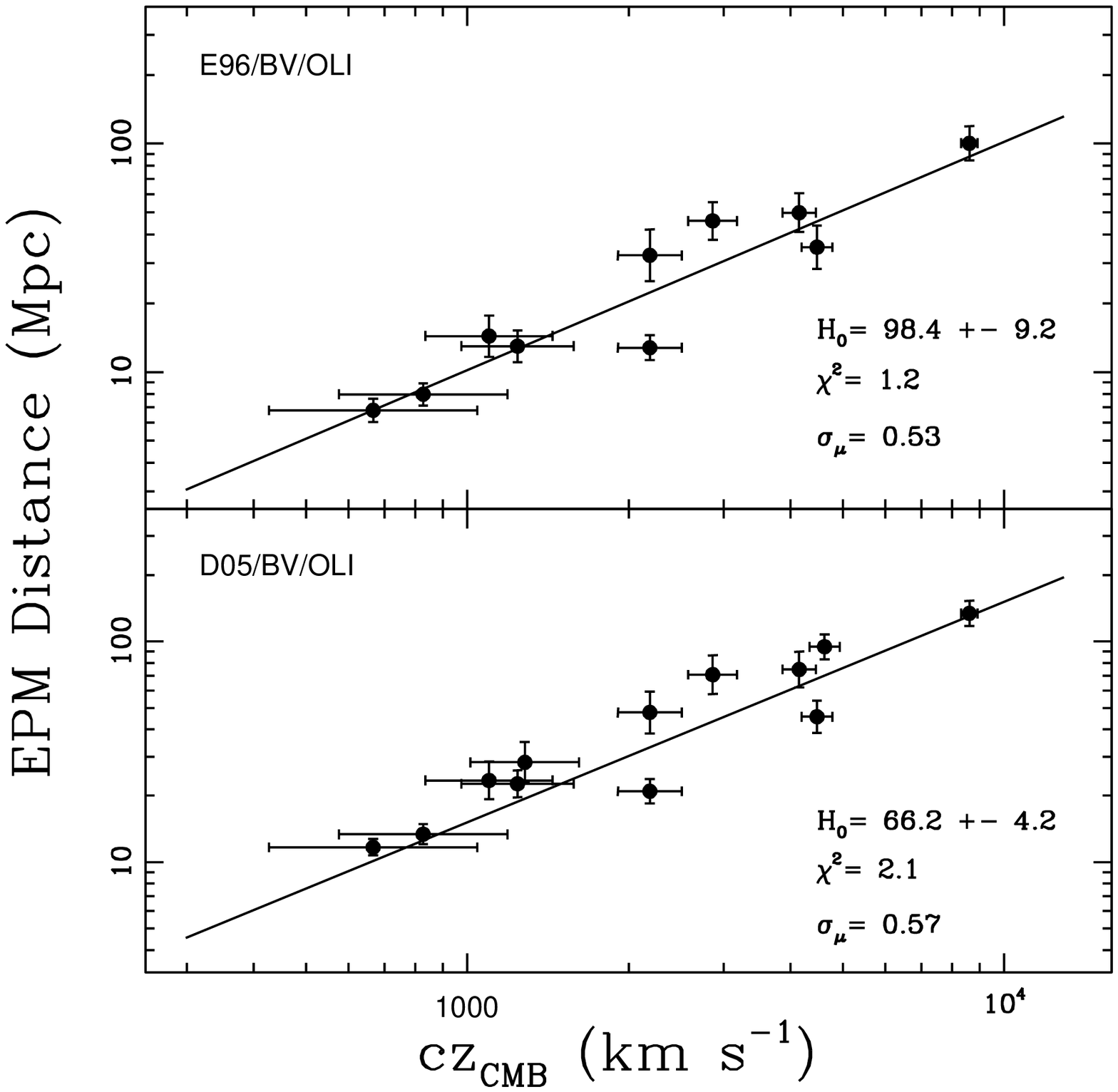}
\caption{Hubble diagram using the \{BV\} filter subset and 
{\rm OLI} reddening. ~\label{fig_bv_OLI.HD}}
\end{figure}

\begin{figure}
\plotone{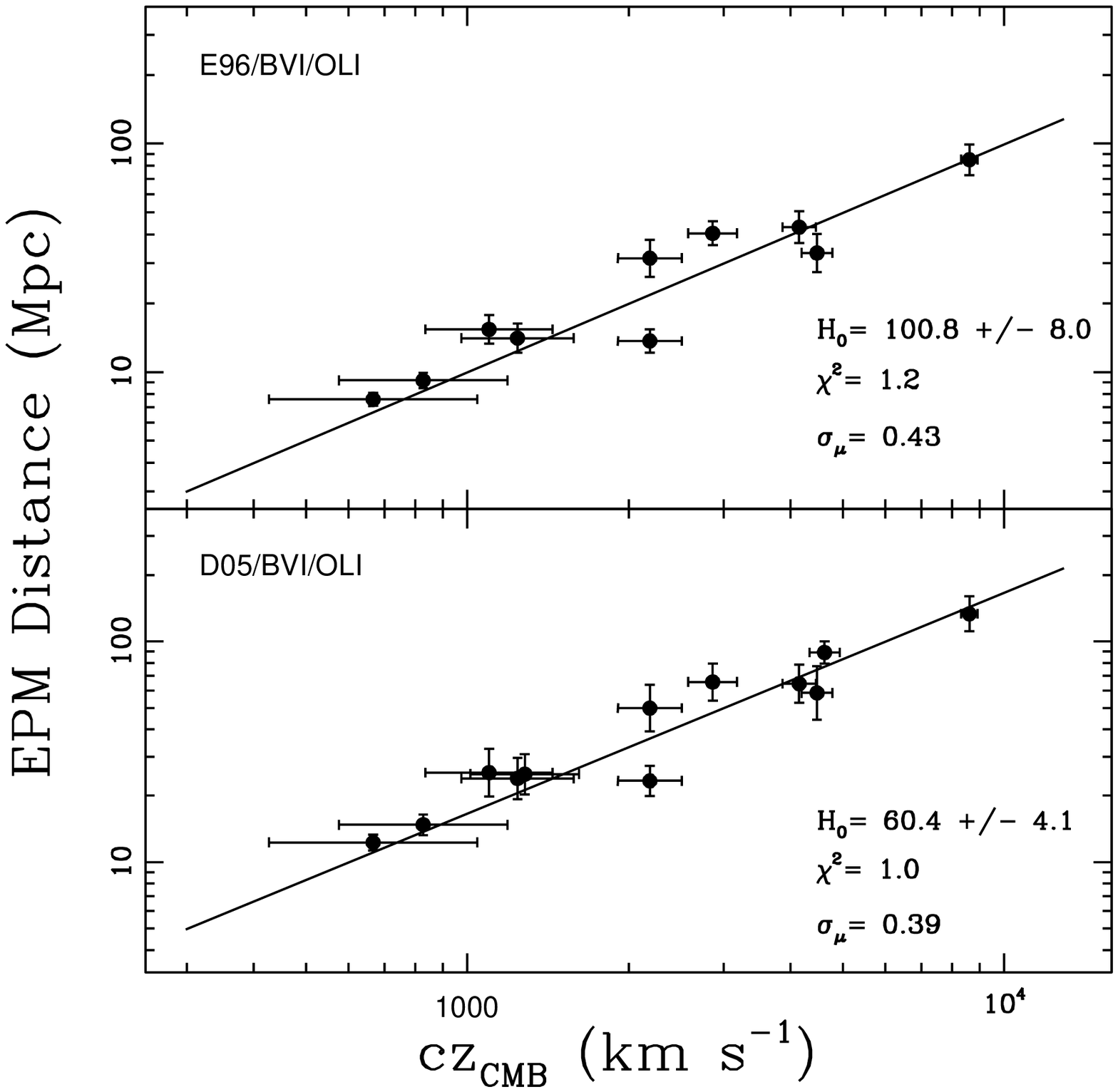}
\caption{Hubble diagram using the \{BVI\} filter subset and 
{\rm OLI} reddening. ~\label{fig_bvi_OLI.HD}}
\end{figure}

\begin{figure}
\plotone{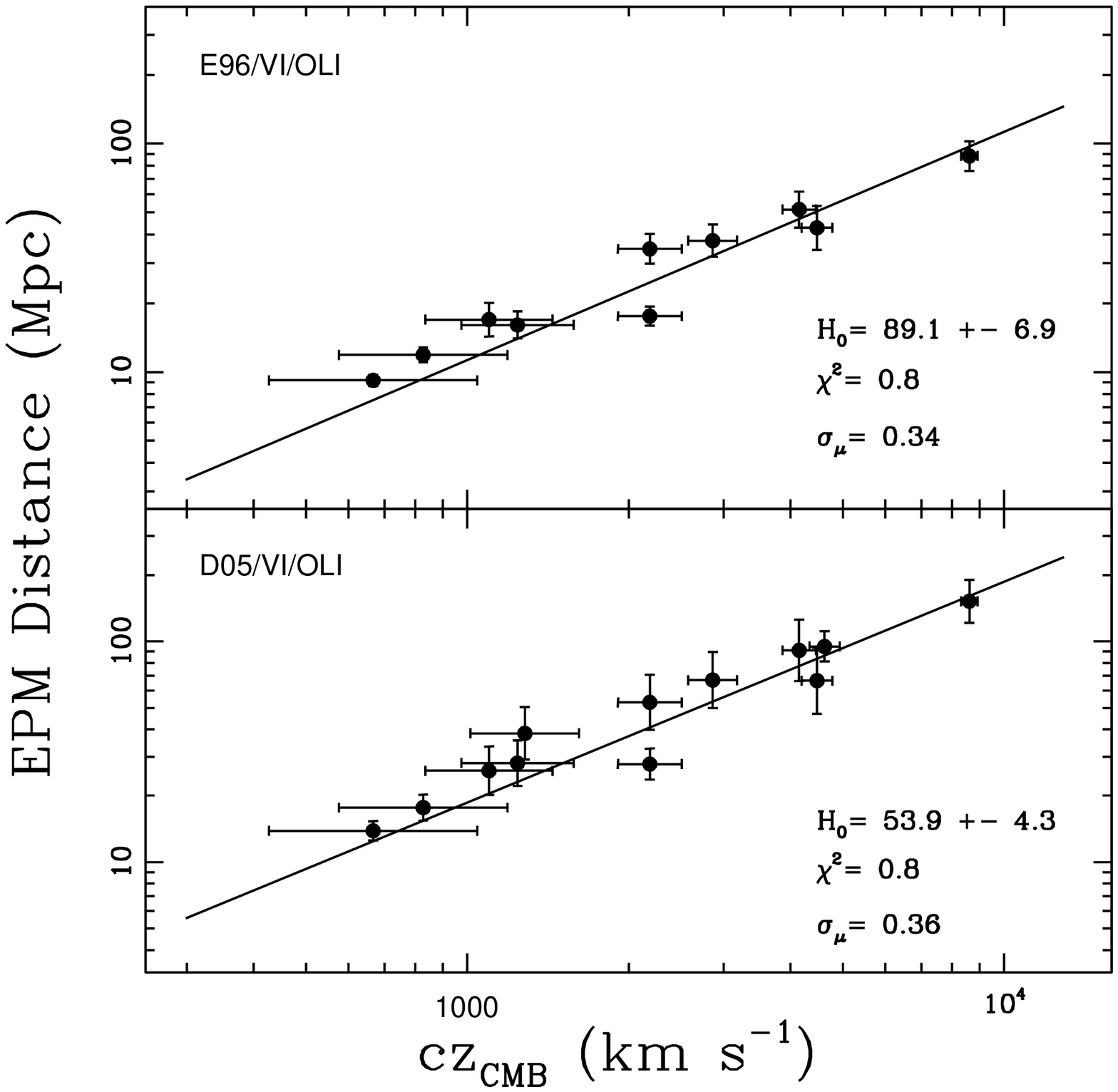}
\caption{Hubble diagram using the \{VI\} filter subset and 
{\rm OLI} reddening. ~\label{fig_vi_OLI.HD}}
\end{figure}

\begin{figure}
\plotone{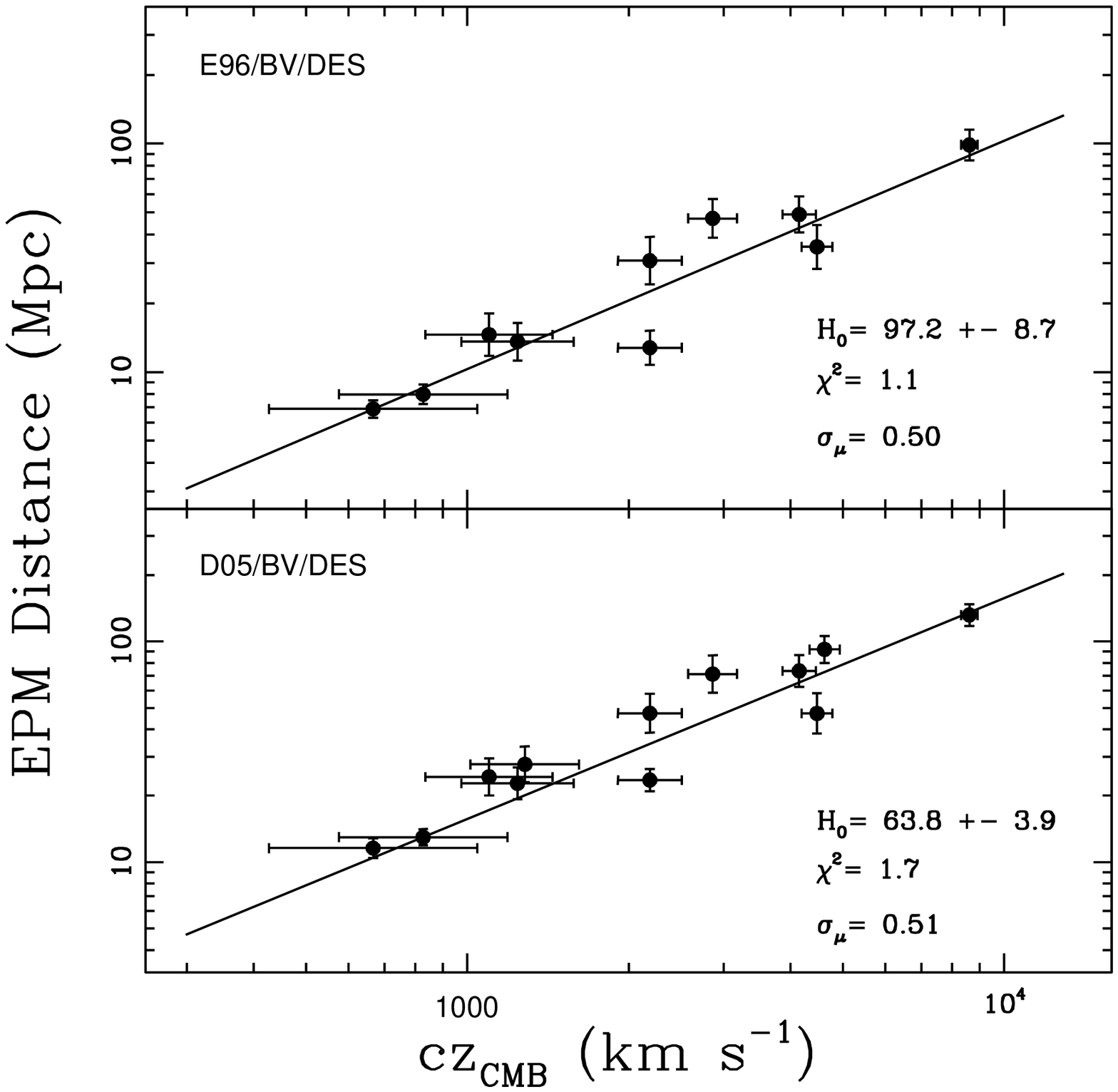}
\caption{Hubble diagram using the \{BV\} filter subset and 
{\rm DES} reddening. ~\label{fig_bv_DES.HD}}
\end{figure}

\begin{figure}
\plotone{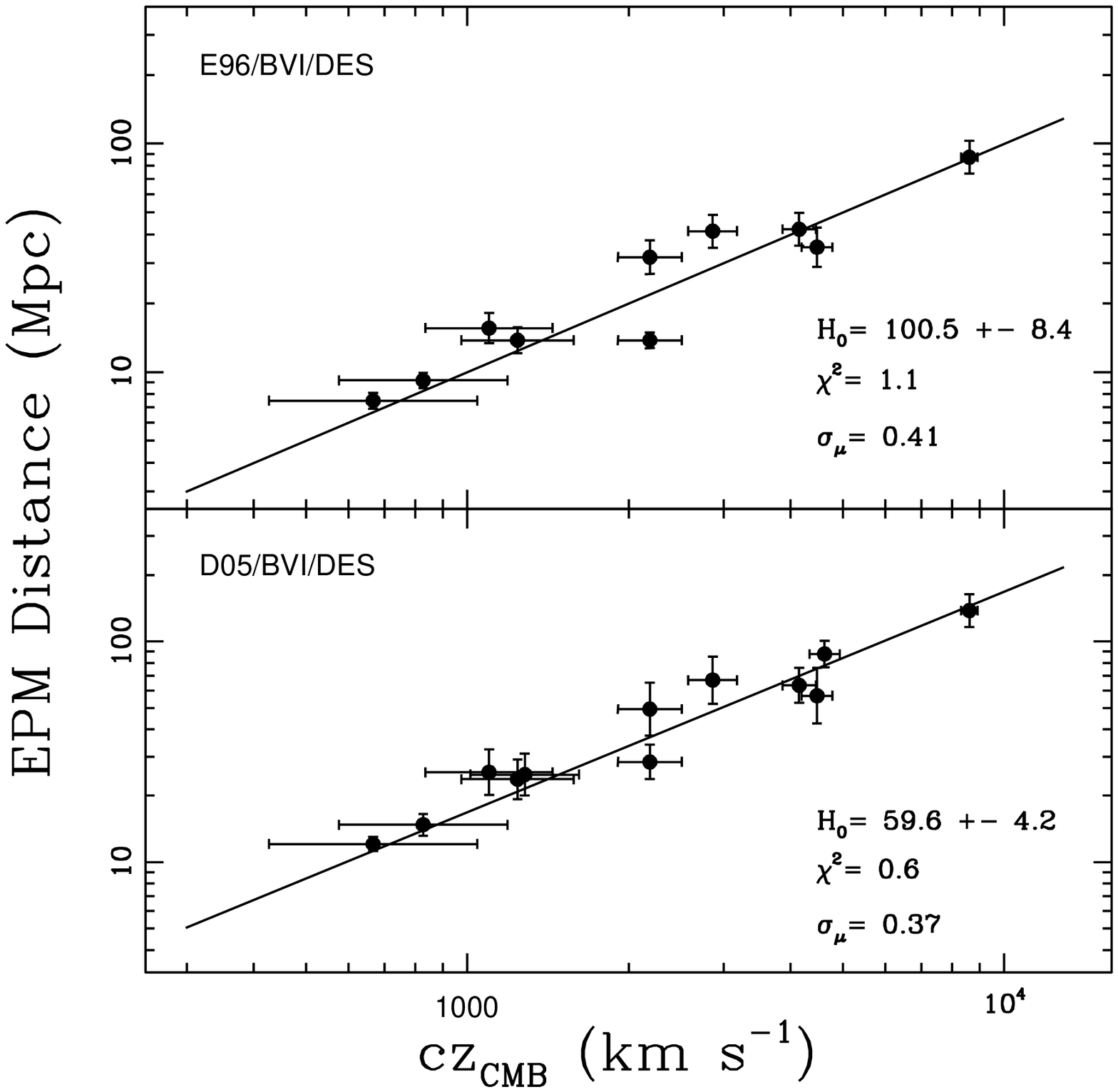}
\caption{Hubble diagram using the \{BVI\} filter subset and 
{\rm DES} reddening. ~\label{fig_bvi_DES.HD}}
\end{figure}

\begin{figure}
\plotone{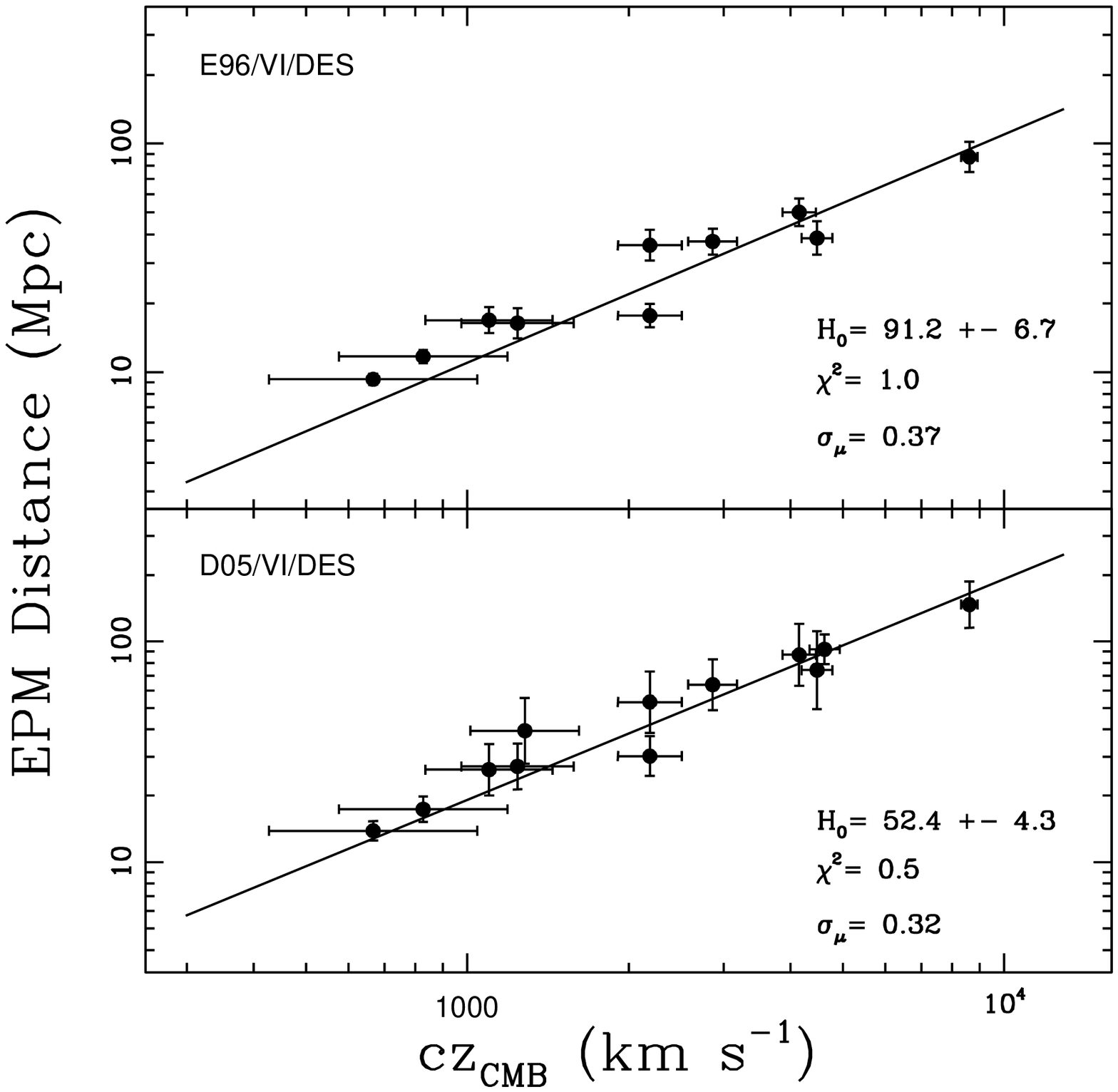}
\caption{Hubble diagram using the \{VI\} filter subset and 
{\rm DES} reddening. ~\label{fig_vi_DES.HD}}
\end{figure}

Since the discovery of the expansion of the Universe \citep{Hub29},
the determination of the expansion rate, the Hubble constant (${\rm
H_{0}}$), has become one of the most important challenges in astronomy
and cosmology. Using the velocity-distance relation (Hubble diagram)
calibrated with the Cepheid period-luminosity relation, \citet{Hub31}
obtained ${\rm H_{0}\sim\,500~km~s^{-1}~Mpc^{-1}}$. During the
second half of the 20th century, the Cepheid relation was
significantly improved, and new Hubble diagrams were obtained,
yielding Hubble constants in the range $\sim\,$50--100 ${\rm km~
s^{-1}~ Mpc^{-1}}$. Today the discrepancy is not over, but there is a
convergence into a value of ${\rm H_{0}\sim\,}$65--80 ${\rm km~ s^{-1}~
Mpc^{-1}}$) \citep{San06,Fre01,Riess05}. \newline \indent
  In this work we applied the EPM method to 12 SNe using two sets of
dilution factors ({\rm E96}, {\rm D05}), two extinction determination
methods ({\rm OLI}, {\rm DES}), and three filter subsets ($\{BV\},\,
\{BVI\}$ and $\{VI\}$) to derive their distances. In order to obtain the
host-galaxy redshifts relative to the cosmic microwave
background (CMB), we corrected the heliocentric host-galaxy redshifts
for the peculiar velocity of the Sun relative to the CMB rest
frame. For this purpose we added a velocity vector of 371 ${\rm km~
s^{-1}}$ in the direction $(l,b)=(264.14^{\circ},48.26^{\circ})$
\citep{Fix96} to the heliocentric redshifts. 
The resulting CMB redshifts are given in Table \ref{tab_SN_list}. 
\newline \indent
  Using the CMB host-galaxy redshifts, we constructed 12 different
Hubble diagrams. In each case we computed a linear fit weighting the error in 
distance and redshift (assumed to be 300 ${\rm km~ s^{-1}}$
for all SNe) in order to derive the Hubble constant.
Figures \ref{fig_bv_OLI.HD}--\ref{fig_vi_OLI.HD}
show the Hubble diagrams obtained with {\rm OLI} reddenings, from the
$\{BV\},\, \{BVI\}$ and $\{VI\}$ filter subsets, respectively. Figures
\ref{fig_bv_DES.HD}--\ref{fig_vi_DES.HD} show the same diagrams but
this time using {\rm DES} extinctions. Each diagram is labeled with
the derived Hubble constant, the reduced $\chi^{2}$, and the dispersion
in distance modulus $\sigma_{\mu}$ from the linear fit. The resulting
${\rm H_0}$ values are summarized in Table \ref{tab_H0_values}. \newline \indent
  There is a systematic difference in the ${\rm H_{0}}$
values obtained with the {\rm E96} and {\rm D05} models. Using {\rm
E96} we obtained ${\rm H_{0}=}$ 89--101 ${\rm km~ s^{-1}~ Mpc^{-1}}$
while {\rm D05} yielded ${\rm H_{0}=}$ 52--66 ${\rm km~ s^{-1}~
Mpc^{-1}}$. This difference arises both from the systematically
higher {\rm D05} dilution factors and the different photospheric 
velocity conversion between both
models. These two effects combined lead to differences of $\sim\,40~\%$ in the
EPM distances. \newline \indent
The use of different filter subsets leads to ${\rm H_{0}}$ values
consistent within ${\rm 1\,\sigma}$ for a fixed atmosphere model. This
is a very important result, because it suggest an internal consistency
for each set of atmosphere models. However, the use of different
filter subsets produces significant differences in dispersion,
increasing from ${\rm \sigma_{\mu}\sim\,0.3}$ (\{VI\}) to ${\rm
\sigma_{\mu}\sim\,0.4}$ (\{BVI\}) and ${\rm \sigma_{\mu}\sim\,
0.5}$ (\{BV\}) (see Table \ref{tab_Dispersion_values}). The special
case of {\rm D05} with \{VI\} and {\rm DES}, leads to ${\rm
\sigma_{\mu}\, =\, 0.32}$, which corresponds to $\sim\,15$\,\%
of error in distance. Clearly when the $B$ band is employed, the
dispersion in the Hubble diagram increases considerably.  
This is due to the fact that the dilution factors in the B band, 
are mostly determined by the effect of line blanketing rather 
than by electron scattering, so the assumption of a ``dilute" Black Body
becomes less reliable. It is also possible that metallicity differences of the SNe
could explain part of this scatter, although both atmosphere model sets 
predict a modest effect of metallicity in the emergent flux at wavelengths 
longer than $\sim$4000 \AA. \newline \indent
  As expected, it can be noted that there are no significant differences
in the ${\rm H_{0}}$ values and in the Hubble diagram dispersion
between the {\rm DES} and {\rm OLI} reddening methods. This is
because there is no systematic difference in the reddening between
both methods (see \mbox{$\S$ 3.5).} However, the {\rm DES} method leads
to somewhat lower dispersion in the Hubble diagrams than the {\rm OLI}
technique. \newline \indent
  Finally, SN 2003hl and SN 2003iq are of particular interest because
they both exploded in the same galaxy. To our disappointment, all 12
possible combinations of filter subsets, reddening, and atmosphere
models lead to significant differences in the EPM distance to the host
galaxy. The most extreme case is the $\{BV\}$, {\rm E96}, and {\rm OLI}
combination, which leads to a distance of 32.5 $\pm$ 8.5 Mpc to SN
2003iq and 12.8 $\pm$ 1.6 Mpc to SN 2003hl (a difference of 2.3
sigma). The smallest discrepancy occurs with the $\{VI\}$, {\rm D05}, 
and {\rm DES} combinations (30.3 $\pm$ 6.3 and 53.3 $\pm$ 17.1 Mpc for
SN 2003hl and SN 2003iq, respectively), which is also the combination
that produces the lowest dispersion in the Hubble diagram. As
discussed in \S 3, the EPM solutions to SN 2003iq yield an explosion
time inconsistent with a pre-discovery image, hence the EPM
distance to SN 2003iq is quite suspicious. 
The light curve and the spectroscopic
velocity evolution of both SNe are typical of the Type IIP class, so this
is not the reason for the inconsistency in the EPM solutions.
Given the high host-galaxy extinction to SN~2003hl, it is conceivable
that the discrepancy could be due to a departure from the standard 
reddening law. We explored this possibility by varying $R_V$ over 
the range 1.1-3.1 using the $\{VI\}$ and {\rm D05} models. For the case
$R_V$=3.1 we obtain D(03hl)/D(03iq)=0.57, while for $R_V$=1.1 we get
D(03hl)/D(03iq)=0.72. This exercise shows that, although the discrepancy
is not completely removed, lowering $R_V$ brings the two distances in better
agreement. Interestingly, \citet{OLI08} found an optimal value of $R_V$=1.4 
from a set of 34 SNe~IIP using the SCM. All this suggests that either 1) the dust
around SNe~IIP has different properties than the Galactic dust, or 2)
SNe~IIP are surrounded by Galactic-like dust whose geometric distribution
is responsible for an abnormal low value of $R_V$ \citep{goobar08}.


\subsection{External Calibration and the Internal precision of the EPM}

\begin{figure}
\plotone{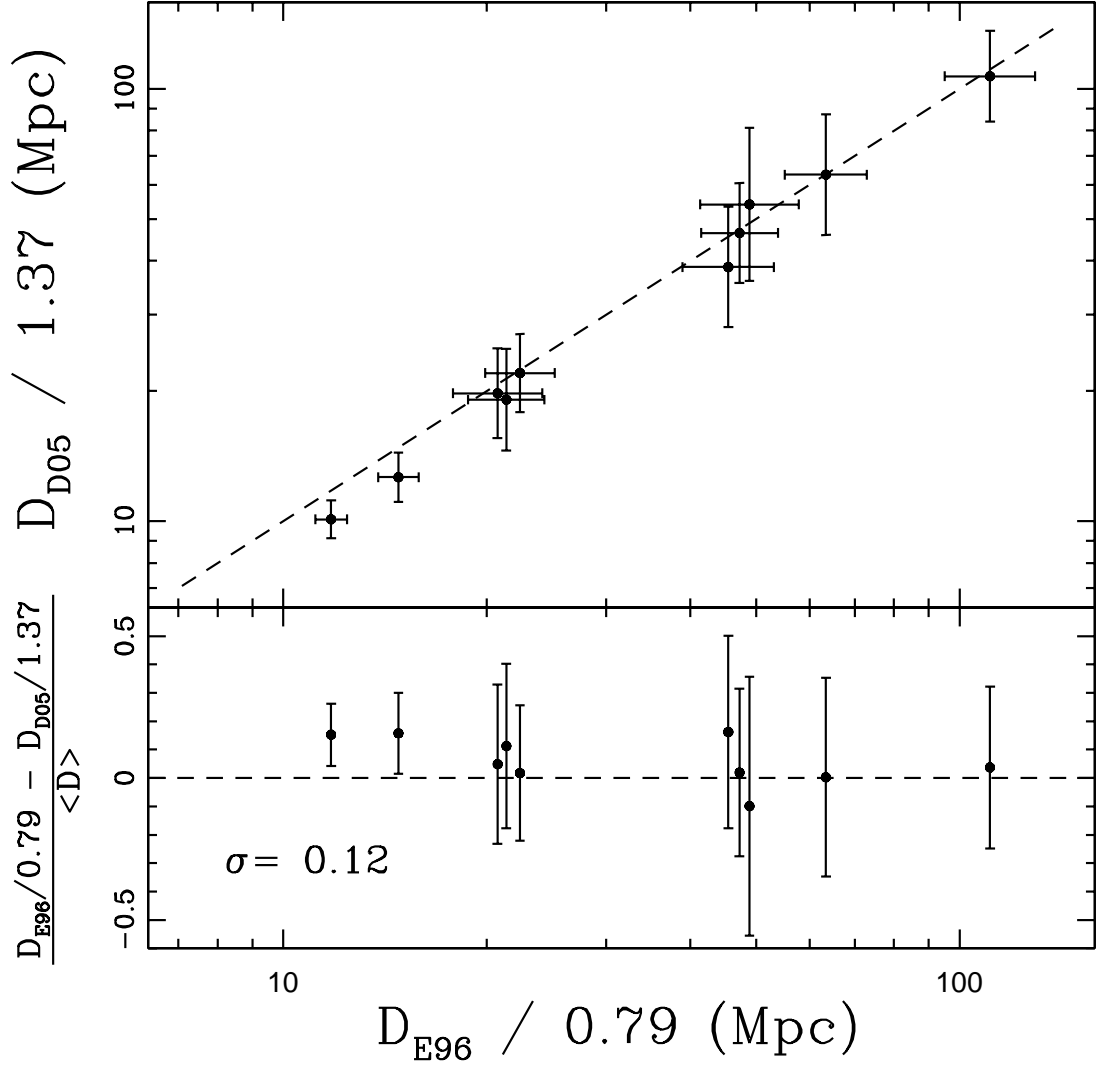}
\caption{Top panel: {\rm D05} distances versus {\rm E96} distances
corrected to the {\it HST Key Project} Cepheid scale. The dashed line
shows slope unity.  Bottom panel: differences between the corrected
distances normalized to the average of the {\rm E96} and {\rm D05}
corrected distances. The 12\% scatter reflects the internal precision
of the EPM.
~\label{fig_Cepheids_vi_DES.HD}}
\end{figure}

In the previous section we have shown that there is a systematic
difference in the ${\rm H_0}$ values derived using the {\rm E96} and
the {\rm D05} models. In order to remove this systematic effect, we
applied a calibration factor (given by the ratio between some external
${\rm H_0}$ value and the EPM ${\rm H_0}$ value) to the distances
derived using {\rm E96} and {\rm D05}. For this purpose we used the
value of ${\rm H_0\,=\,72~ km~ s^{-1}~ Mpc^{-1}}$ derived from the
{\it HST Key Project} \citep{Fre01}. This external calibration allows
us to bring the EPM distances to the Cepheid scale and allows us to
remove the systematic difference in the EPM distances between {\rm
E96} and {\rm D05}. Figure \ref{fig_Cepheids_vi_DES.HD} shows (top
panel) the {\rm D05} distances versus the {\rm E96} distances divided
by a calibration factor of 1.37 and 0.79, respectively. In both cases
the EPM distances were derived using the $\{VI\}$ filter subset and
the {\rm DES} reddening. As can be seen, after applying this
correction, the systematic differences disappear. The dashed line in
the top panel corresponds to the one to one relation. Also, in Figure
\ref{fig_Cepheids_vi_DES.HD} (bottom panel) are plotted the
differences between the corrected {\rm E96} and {\rm D05} distances,
normalized to the corresponding average between the corrected {\rm
E96} and {\rm D05} distance. We found a standard deviation of ${\rm
\sigma\,=\,0.12}$. Since the dispersion arises from the combined
errors in the {\rm E96} and the {\rm D05} distances, the internal
random errors in any of the EPM implementation must be less than 12\%.
Note that this scatter is smaller than the $\sim\,15$\,\% dispersion seen
in the Hubble diagrams, which is affected by the peculiar motion of
the host galaxies. The 12\% scatter is independent of the redshift and
must be an upper value of the internal precision of the EPM.

\section{Conclusions}

In this work we have applied the EPM method to 12 SNe IIP. We
contructed 12 different Hubble diagrams, using three different filter
subsets ($\{BV\},\,\{BVI\},\,\{VI\}$), two atmosphere models ({\rm E96,
D05}), and two methods to determine the amount of host-galaxy
extinction ({\rm DES, OLI}). Our main conclusions are as follows. 
\newline \indent
The EPM must be restricted to the first $\sim\,$45--50 days from
explosion. After that epoch the method may display departures from
linearity in the $\theta/v$ versus time relation, and therefore an
internal inconsistency.
\newline \indent
The results are less precise when the $B$ band is used in the EPM
analysis, regardless of the atmosphere models employed ({\rm E96} or
{\rm D05}). The dispersion in the Hubble diagrams increases
considerably from 0.3 to 0.5 mag when the $B$ band is included and the
$V$ band is removed from the filter subset. Despite the loss in
precision, there are no significant differences in the resulting
distances.
\newline \indent
We investigated the effect of host-galaxy reddening in the EPM
distances. For this purpose we computed many EPM solutions varying the
amount of visual extinction, and we found that a difference of $\Delta
A_V = 0.5$ mag leads on average to a difference of $\sim\,$5--10\,\% in
distance. Therefore, we conclude that the method is quite insensitive
to the effect of dust.
\newline \indent
Systematic differences in the atmosphere models lead to
$\sim\,50$\,\% differences in the EPM distances and to values of
${\rm H_0}$ between 52 and 101 ${\rm km~ s^{-1}~ Mpc^{-1}}$. This
effect is due to the systematic difference in the photospheric
velocity conversion and the dilution factors. The latter is currently
the greatest source of uncertainty in the EPM method. \newline \indent  
The Hubble diagram with the lowest dispersion (${\rm
\sigma_{\mu}\,=\,0.32}$ mag) was obtained using the combination {\rm
D05}, $\{VI\}$, {\rm DES}. Despite the systematic uncertainties in the
EPM, this dispersion is quite low and corresponds to a precision of
$\sim\,15$\,\% in distance. This precision is similar to that of the SCM
method for type II SNe \citep{Ham02,OLI08} and to the Tully-Fisher
relation for spiral galaxies with a dispersion of ${\rm \sigma \sim\,
0.30}$ mag \citep{Sak00}. However, the EPM dispersion is greater than
that of the M/$\Delta m_{15}$ relation for Type Ia SNe, which has a
dispersion of $\sigma \sim\,$0.15--0.20 mag; however, if the EPM is
applied to a sample of SNe~IIP in the Hubble flow, the dispersion in
the Hubble diagram might decrease. \newline \indent 
Finally, despite the systematic differences in the ${\rm H_0}$ value,
EPM has great potential as an extragalactic distance indicator and can
potentially be applied to a sample of high-redshift SNe IIP in order
to check in an independent way the accelerating expansion of the
universe.
         
\subsection*{Acknowledgments}

We thank Luc Dessart and Ronald Eastman for providing us with their
atmosphere models. 
We are also grateful to Robert Kirshner for providing some spectra used in this work. 
We acknowledge support from the Millennium Center for Supernova Science 
through grant P06-045-F funded by ``Programa Bicentenario de Ciencia y 
Tecnolog\'ia de CONICYT'' and ``Programa Iniciativa Cient\'ifica Milenio de MIDEPLAN''.
Additional support was provided by CONICYT through Centro de Astrof\'isica FONDAP 
(grant 15010003), Center of Excellence in Astrophysics and Associated Technologies 
(grant PFB 06), and Fondecyt (grant 1060808).
A.V.F.'s supernova group at UC Berkeley and KAIT at Lick Observatory are supported by NSF
grant AST--06074845, as well as by generous gifts from the Sylvia and
Jim Katzman Foundation and the TABASGO Foundation.


\clearpage

\clearpage

\clearpage

\clearpage

\clearpage

\clearpage

\clearpage

\clearpage

\clearpage

\clearpage

\clearpage

\clearpage

\clearpage

\clearpage

\clearpage

\clearpage

\clearpage

\clearpage

\clearpage

\clearpage

\clearpage

\clearpage

\clearpage

\clearpage

\clearpage

\clearpage

\clearpage

\clearpage

\clearpage

\clearpage

\clearpage



\clearpage
\begin{deluxetable}{ccccc}
\tablewidth{0pt}
\tablecaption{Heliocentric and CMB redshifts for the SNe used in this work. 
\label{tab_SN_list}}
\tablehead{
\colhead{SN} & \colhead{Host Galaxy} & \colhead{$cz_{helio}$} & 
\colhead{source\tablenotemark{a}} & \colhead{$cz_{CMB}$} \\
\colhead{}  & \colhead{} & \colhead{${\rm (km~ s^{-1})}$} & 
\colhead{} &  \colhead{${\rm (km~ s^{-1})}$} }
\startdata
1992ba &  NGC 2082                &  1092  & here & 1245 \\
1999br &  NGC 4900                &  960   & NED  & 1285 \\  
1999em &  NGC 1637                &  800   & L02  & 670  \\ 
1999gi &  NGC 3184                &  543   & here & 831  \\ 
2002gw &  NGC 0922                &  3117  & here & 2877 \\ 
2003T  &  UGC 04864               &  8368  & NED  & 8662 \\ 
2003bl &  NGC 5374                &  4382  & NED  & 4652 \\ 
2003bn &  2MASX J10023529-2110531 &  3829  & NED  & 4173 \\ 
2003ef &  NGC 4708                &  4440  & here & 4503 \\ 
2003hl &  NGC 0772                &  2265  & here & 2198 \\ 
2003hn &  NGC 1448                &  1347  & here & 1102 \\ 
2003iq &  NGC 0772                &  2364  & here & 2198 \\ 
\enddata
\tablenotetext{a}{\begin{small} The NED values correspond to the
redshifts of the host-galaxy nucleus, while the values measured in this
work (``here") were measured from narrow emission lines of H~II regions
at the SN position. Also, L02 corresponds to the value adopted from
\citet{Leo02b}.\end{small}}
\end{deluxetable}

\clearpage
\begin{deluxetable}{cccccccccccc}
\tablewidth{0pt}
\tablecaption{Dilution factor coefficients and dispersion \label{tab_zeta_coef}} 
\tablehead{
\multicolumn{3}{c}{} & \multicolumn{2}{c}{{\rm E96}} & \multicolumn{3}{c}{} & 
\multicolumn{2}{c}{{\rm D05}} \\
\cline{3-6} \cline{8-11}    
\colhead{Filter subset} & \colhead{} & \colhead{$b_0$} & \colhead{$b_1$} & \colhead{$b_2$} &  
\colhead{$\sigma$} & \colhead{} & \colhead{$b_0$} & \colhead{$b_1$} 
& \colhead{$b_2$} & \colhead{$\sigma$} }
\startdata
\{BV\}  & & 0.756 & -0.900 & 0.520 & 0.048 & & 0.593 & -0.450 & 0.403 & 0.075 \\
\{BVI\} & & 0.733 & -0.693 & 0.373 & 0.027 & & 0.711 & -0.476 & 0.308 & 0.068 \\
\{VI\}  & & 0.702 & -0.531 & 0.265 & 0.029 & & 0.915 & -0.747 & 0.371 & 0.077 \\
\enddata
\end{deluxetable}

\clearpage
\begin{deluxetable}{ccc}
\tablewidth{0pt}
\tablecaption{H$\beta$ to photospheric velocity ratio coefficients and dispersion.
\label{tab_ratio_coef}}
\tablehead{
\colhead{$j$}  & \colhead{$a_{j}({\rm E96})$} & \colhead{$a_{j}({\rm D05})$}}
\startdata
0 & 1.775 & 1.014       \\
1 & $-1.435 \times 10^{-4}$ & $4.764 \times 10^{-6}$    \\
2 & $6.523 \times 10^{-9}$ & $-7.015 \times 10^{-10}$  \\
$\sigma$ & 0.06 & 0.04 \\
\enddata
\end{deluxetable}

\clearpage
\begin{deluxetable}{cccc}
\tablewidth{0pt}
\tablecaption{SN host galaxy and Galactic extinction adopted. 
\label{tab_reddening}}
\tablehead{
\colhead{SN}  & \colhead{${\rm A_{V}}$ ({\rm OLI})}\tablenotemark{a}
& \colhead{${\rm A_{V}}$ ({\rm DES})} \tablenotemark{b}
& \colhead{${\rm A_{V}}$ (IR maps)} \tablenotemark{c} \\
\colhead{}  & \colhead{{\rm $Host$}}  & \colhead{${\rm Host}$}
& \colhead{${\rm Galactic}$}
}
\startdata
1992ba & 0.30 (0.15)  &  0.43 (0.16)  &  0.193 (0.031) \\
1999br & 0.94 (0.20)  &  0.25 (0.16)  &  0.078 (0.012) \\
1999em & 0.24 (0.14)  &  0.31 (0.16)  &  0.134 (0.021) \\
1999gi & 1.02 (0.15)  &  0.56 (0.16)  &  0.055 (0.009) \\
2002gw & 0.18 (0.16)  &  0.40 (0.19)  &  0.065 (0.010) \\
2003T  & 0.35 (0.15)  &  0.53 (0.31)  &  0.104 (0.017) \\
2003bl & 0.26 (0.15)  &  0.00 (0.16)  &  0.090 (0.014) \\
2003bn & -0.04 (0.15) &  0.09 (0.16)  &  0.215 (0.034) \\
2003ef & 0.98 (0.15)  &  1.24 (0.25)  &  0.153 (0.024) \\
2003hl & 1.72 (0.18)  &  1.24 (0.25)  &  0.241 (0.039) \\
2003hn & 0.46 (0.14)  &  0.59 (0.25)  &  0.047 (0.008) \\
2003iq & 0.25 (0.16)  &  0.37 (0.16)  &  0.241 (0.039) \\
\enddata
\tablenotetext{a}{\begin{small}  \citet{OLI08}.
\end{small}}
\tablenotetext{b}{\begin{small}  \citet{D06,D08a}.
\end{small}}
\tablenotetext{c}{\begin{small} \citet{Sch98}.
\end{small}}
\end{deluxetable}

\clearpage
\begin{deluxetable}{ccccc}
\tablewidth{0pt}
\tablecaption{EPM distances. 
\label{tab_VI_DES.EPM}}
\tablehead{
\colhead{SN}  & \colhead{$D_{(E96)}$} & \colhead{$t_{0(E96)}$} &
\colhead{$D_{(D05)}$} & \colhead{$t_{0(D05)}$} \\
\colhead{}  & \colhead{(Mpc)} & \colhead{(JD--2448000)} &
\colhead{(Mpc)} & \colhead{(JD--2448000)}}
\startdata
1992ba & 16.4 (2.5)   & 883.9 (3.0)  & 27.2 (6.5)   & 879.8 (5.6)  \\
1999br & \nodata      & \nodata      & 39.5 (13.5)  & 3275.6 (7.7)  \\
1999em & 9.3 (0.5)    & 3476.3 (1.1) & 13.9 (1.4)   & 3474.0 (2.0)  \\
1999gi & 11.7 (0.8)   & 3517.0 (1.2) & 17.4 (2.3)   & 3515.6 (2.4)  \\
2002gw & 37.4 (4.9)   & 4557.9 (2.7) & 63.9 (17.0)  & 4551.7 (7.6)  \\
2003T  & 87.8 (13.5)  & 4654.2 (2.7) & 147.3 (35.7) & 4648.9 (6.1)  \\
2003bl & \nodata      & \nodata      & 92.4 (14.2)  & 4694.5 (2.0)  \\
2003bn & 50.2 (7.0)   & 4693.4 (2.7) & 87.2 (28.0)  & 4687.0 (9.0)  \\
2003ef & 38.7 (6.5)   & 4759.8 (4.7) & 74.4 (30.3)  & 4748.4 (15.6) \\
2003hl & 17.7 (2.1)   & 4872.3 (1.7) & 30.3 (6.3)   & 4865.4 (5.9)  \\
2003hn & 16.9 (2.2)   & 4859.5 (3.8) & 26.3 (7.1)   & 4853.8 (9.3)  \\
2003iq & 36.0 (5.6)   & 4909.6 (4.3) & 53.3 (17.1)  & 4905.6 (9.5)  \\
\enddata
\tablecomments{The distances were derived using the $\{VI\}$ filter 
subset and {\rm DES} reddening.}
\end{deluxetable}

\clearpage
\begin{deluxetable}{cc}
\tablewidth{0pt}
\tablecaption{Error Sources \label{tab_errors}}
\tablehead {
\colhead{Error Source}  & \colhead{Typical Error} }
\startdata
Photometry & 0.02 {\rm mag}\\
SN redshift & 50 / 200 (${\rm km~ s^{-1}}$) \tablenotemark{a}\\
Foreground extinction & 0.02 {\rm mag} \\
Host galaxy extinction & 0.15 {\rm mag}\\
Line expansion velocity & 85 (${\rm km~ s^{-1}}$) \\
Photospheric velocity conversion & 0.06 / 0.04 \tablenotemark{b}\\
Dilution Factors & 0.03 / 0.07 \tablenotemark{b}\\
\enddata
\tablenotetext{a}{\begin{small} Corresponds to the redshifts measured
in this work and those taken from NED, respectively.
\end{small}}
\tablenotetext{b}{\begin{small} Corresponds to the {\rm E96} and {\rm D05} models, respectively. 
\end{small}}
\end{deluxetable}

\clearpage
\begin{deluxetable}{cccccc}
\tablewidth{0pt}
\tablecaption{SN 1999em EPM quantities. 
\label{tab_SN99em_EPM}}
\tablehead {
\colhead{JD-}  & \colhead{$T_{VI}$} & \colhead{$\theta$$\zeta_{VI}$} &
\colhead{$\zeta_{VI}$} & \colhead{$v_{phot}$} & \colhead{$\theta/vel$} \\ 
\colhead{2451000}  & \colhead{(K)} & \colhead{${\rm (10^{15}~ cm~ Mpc^{-1})}$} &
\colhead{} & \colhead{${\rm (km~ s^{-1})}$} & \colhead{${\rm (100~ s~ Mpc^{-1})}$} }
\startdata
482.8 & 14588 (469) & 0.0321 (0.0010) & 0.574 & 11022 & 506.7 (73.0)\\
483.8 & 14349 (462) & 0.0331 (0.0011) & 0.572 & 10355 & 559.6 (81.0)\\ 
484.8 & 13986 (382) & 0.0341 (0.0009) & 0.568 & 9867  & 608.3 (88.1)\\ 
485.2 & 13810 (415) & 0.0345 (0.0011) & 0.566 & 9117  & 669.5 (97.7)\\ 
485.7 & 13550 (414) & 0.0352 (0.0011) & 0.563 & 8942  & 699.7 (102.6)\\ 
485.7 & 13544 (414) & 0.0352 (0.0011) & 0.563 & 8915  & 702.1 (103.0)\\ 
485.8 & 13479 (456) & 0.0353 (0.0012) & 0.562 & 9311  & 675.0 (99.7)\\ 
486.8 & 12812 (425) & 0.0373 (0.0013) & 0.555 & 8817  & 762.6 (113.9)\\ 
487.9 & 11985 (333) & 0.0403 (0.0013) & 0.547 & 8584  & 857.5 (128.9)\\ 
488.8 & 11587 (310) & 0.0413 (0.0013) & 0.544 & 8598  & 882.8 (133.3)\\ 
489.8 & 11352 (256) & 0.0424 (0.0011) & 0.542 & 8476  & 921.1 (138.7)\\ 
491.1 & 11077 (350) & 0.0443 (0.0016) & 0.541 & 7870  & 1040.1 (159.5)\\ 
491.2 & 11055 (358) & 0.0444 (0.0017) & 0.541 & 7824  & 1050.6 (161.4)\\ 
491.7 & 10939 (372) & 0.0453 (0.0018) & 0.540 & 7964  & 1053.3 (162.6)\\ 
492.1 & 10840 (349) & 0.0460 (0.0018) & 0.539 & 7863  & 1083.7 (166.9)\\ 
496.2 & 10264 (312) & 0.0495 (0.0019) & 0.537 & 7031  & 1311.6 (202.6)\\ 
496.7 & 10224 (301) & 0.0497 (0.0018) & 0.537 & 7172  & 1290.7 (199.0)\\ 
501.2 & 9610 (224)  & 0.0526 (0.0016) & 0.537 & 5921  & 1653.2 (252.9)\\ 
501.7 & 9386 (185)  & 0.0548 (0.0014) & 0.538 & 6107  & 1667.6 (253.4)\\ 
501.7 & 9384 (185)  & 0.0548 (0.0014) & 0.538 & 6250  & 1630.3 (247.7)\\ 
501.8 & 9362 (189)  & 0.0551 (0.0015) & 0.538 & 6506  & 1572.5 (238.9)\\ 
504.8 & 8907 (173)  & 0.0589 (0.0016) & 0.542 & 5991  & 1813.0 (274.0)\\ 
506.8 & 8655 (162)  & 0.0605 (0.0016) & 0.545 & 5691  & 1950.5 (293.2)\\ 
510.8 & 8248 (63)   & 0.0649 (0.0007) & 0.553 & 5156  & 2276.6 (334.0)\\ 
514.8 & 7819 (92)   & 0.0705 (0.0013) & 0.565 & 4877  & 2557.8 (370.3)\\ 
\enddata
\tablecomments{The EPM quantities were derived using the \{VI\} filter
subset, the {\rm DES} reddening and the {\rm D05} models.}
\end{deluxetable}

\clearpage
\begin{deluxetable}{ccccc}
\tablewidth{0pt}
\tablecaption{Summary of H$_0$ values. \label{tab_H0_values}}
\tablehead {
\colhead{} & \colhead{} & \colhead{$\{BV\}$} & \colhead{$\{BVI\}$} & \colhead{$\{VI\}$} }
\startdata
{\rm E96/OLI} & & 98.4 (9.2) & 100.8 (8.0) & 89.1 (6.9) \\
{\rm E96/DES} & & 97.2 (8.7) & 100.5 (8.4) & 91.2 (6.7) \\
{\rm D05/OLI} & & 66.2 (4.2) & 60.4 (4.1)  & 53.9 (4.3) \\
{\rm D05/DES} & & 63.8 (3.9) & 59.6 (4.2)  & 52.4 (4.3) \\
\enddata
\end{deluxetable}

\clearpage
\begin{deluxetable}{ccccc}
\tablewidth{0pt}
\tablecaption{Summary of dispersions in Hubble diagrams. \label{tab_Dispersion_values}}
\tablehead {
\colhead{} & \colhead{} & \colhead{$\{BV\}$} & \colhead{$\{BVI\}$} & \colhead{$\{VI\}$} }
\startdata
{\rm E96/OLI} & & 0.53  & 0.43 & 0.34 \\
{\rm E96/DES} & & 0.50  & 0.41 & 0.37 \\
{\rm D05/OLI} & & 0.57  & 0.39 & 0.36 \\
{\rm D05/DES} & & 0.51  & 0.37 & 0.32 \\
\enddata
\end{deluxetable}


\end{document}